\documentclass[reprint,showpacs,amsmath,amssymb,floatfix,nofootinbib]{revtex4-1}
\usepackage{graphicx,bm,dcolumn,mathtools,times}

\usepackage{multirow}

\newcommand{\llas}[0]{\langle\langle}
\newcommand{\rras}[0]{\rangle\rangle}
\newcommand{\nag}{{\phantom{\dag}}}
\newcommand{\lso}[0]{\lambda}
\newcommand{\rmi}{\text{i}}

\newcommand{\chicl}{\chi}
\newcommand{\C}{C}
\newcommand{\DOF}{\text{DOF}}
\newcommand{\Rxi}{R_\xi}
\newcommand{\Rxii}[1]{R^{(#1)}_\xi}
\newcommand{\Rxis}[2]{R_{\xi,s,#1,#2}}
\newcommand{\Rxist}{\Rxis{1/3}{1/3}}
\newcommand{\Rxish}{\Rxis{1/2}{1/2}}
\newcommand{\Rxishq}{\Rxis{1/2}{1/4}}

\newcolumntype{.}{D{.}{.}{-1}}

\begin{document}

\title{Fermionic quantum criticality in honeycomb and $\pi$-flux Hubbard models:\\ Finite-size scaling of renormalization-group-invariant observables from quantum Monte Carlo}

\author{\firstname{Francesco} \surname{Parisen Toldin}}
\author{\firstname{Martin} \surname{Hohenadler}}
\author{\firstname{Fakher F.} \surname{Assaad}}
\affiliation{\mbox{Institut f\"ur Theoretische Physik und Astrophysik, Universit\"at W\"urzburg, Am Hubland, D-97074 W\"urzburg, Germany}}
\author{\firstname{Igor F.} \surname{Herbut}}
\affiliation{Department of Physics, Simon Fraser University, Burnaby, British Columbia V5A 1S6, Canada}

\begin{abstract}
  We numerically investigate the critical behavior of the Hubbard model on the
  honeycomb and the $\pi$-flux lattice, which exhibits
  a direct transition from a Dirac semimetal to an
  antiferromagnetically ordered Mott insulator. We use
  projective auxiliary-field quantum Monte Carlo simulations and a careful
  finite-size scaling analysis that exploits approximately improved renormalization-group-invariant
  observables. This approach, which is successfully verified for the
  three-dimensional XY transition of the Kane-Mele-Hubbard model, allows us to extract
  estimates for the critical couplings and the critical exponents. The
  results confirm  that the critical behavior for the semimetal to Mott
  insulator transition in the Hubbard model belongs to the
  Gross-Neveu-Heisenberg universality class on both lattices.
\end{abstract}



\maketitle

\section{Introduction}\label{sec:intro}

Understanding quantum phase transitions in which the order parameter
couples to gapless fermions is an old and notorious problem in
condensed matter theory \cite{sachdev}. In spite of recent advances (see,
e.g., Refs.~\cite{PhysRevB.82.075127,PhysRevB.82.075128}),
the transitions in electronic systems with a full Fermi
surface often elude  controlled theoretical approaches. It is therefore
useful to study simpler cases, in which gapless fermionic excitations would
reside near surfaces in reciprocal space with co-dimensions larger than
unity. Aside from providing a fundamentally new universality class (UC) outside of
the usual bosonic $\phi^4$ paradigm, theories with gapless fermions
close to, for example,  Dirac or parabolic points also describe physical
systems of great current interest, such as graphene \cite{Herbut06}, $d$-wave
superconductors \cite{PhysRevLett.85.4940}, or three-dimensional gapless semiconductors
\cite{PhysRevLett.111.206401,PhysRevLett.113.106401} such as gray tin, for instance.  Their detailed
understanding could be the stepping stone towards a more comprehensive
picture of quantum phase transitions in which fermions play a decisive
role in the critical behavior.

The aim of this paper is to investigate in detail fermionic criticality in
lattice models where the kinetic energy provides a regularization of
  the Dirac Hamiltonian. In particular, we consider
the Hubbard model on the honeycomb
\cite{Sorella92,Paiva05,Meng10,PhysRevB.83.024408,*PhysRevB.85.045123,Sorella12,Assaad13} and the $\pi$-flux lattice
\cite{Chen2012,IAS-14}. In the absence of interactions, both
lattice models have the same continuum limit given by four-component Dirac
fermions per spin projection.  At half-filling, the density of states is
proportional to the excitation energy, and the semimetal is therefore
stable against weak interactions.  At strong coupling, both models map onto a
Heisenberg Hamiltonian on a nonfrustrated lattice so that we expect an
antiferromagnetic insulating state.  The transition from the semimetal to the
antiferromagnetic Mott insulator has attracted considerable interest.
Starting from the weak-coupling Dirac Hamiltonian, it is natural to understand
the mass generation as the signature of broken sublattice symmetry triggered
by the antiferromagnetic order \cite{Sorella12,Assaad13}.  In this case, the
critical behavior is naturally described
in terms of Gross-Neveu-Yukawa theory where the broken symmetry is at the
origin of mass generation \cite{Herbut09a}.  In fact, at the mean-field
level, mass generation can occur only as a result of symmetry breaking
\cite{Ryu09}.  Starting from strong coupling, and since the transition occurs
at intermediate values of the Hubbard interaction, one can follow the idea
that dynamically generated higher-order ring-exchange spin processes are able
to frustrate the magnetic order without closing the charge gap
\cite{Schmidt12}.  This scenario implies an intermediate, rotationally
invariant, spin-disordered, insulating phase as proposed in Refs.~\cite{Meng10,PhysRevB.83.024408,*PhysRevB.85.045123,Chen2012}.

Here, we show that a consistent and unbiased understanding of
the transition is obtained by assuming a direct transition from the
semimetal to the Mott insulating phase, as described by Gross-Neveu-Yukawa
theory with $N_f=2$ massless four-component Dirac fermions. In the present case, the corresponding critical behavior belongs to
the so-called Gross-Neveu-Heisenberg UC, where the term
Heisenberg emphasizes the SU(2) symmetry group of the order-parameter field.
Within Gross-Neveu-Yukawa theory, a different number of flavors $N_f$ as well as
other symmetry groups are possible \cite{Herbut09a}.
In this context, the case of $N_f=1$ with Ising $\mathbb Z_2$ symmetry
has been recently investigated in Refs.~\cite{WCT-14,Yao2014} in terms of spinless
fermions on the honeycomb lattice, while the case $N_f=2$ with SU(2) symmetry has been studied in 
Ref.~\cite{PhysRevD.88.021701} by directly simulating the field theory on a lattice.
Here and in the following, we restrict ourselves to the case of $N_f=2$, which is relevant for
the physics of graphene.
From the perspective of Gross-Neveu-Yukawa theory with the Heisenberg SU(2) symmetry,
both the honeycomb and the $\pi$-flux Hubbard lattice models are different
regularizations of the same continuum theory.  Hence, both models should have
the same critical exponents.  Our analysis of the transition is based on the notion of
improved renormalization-group- (RG-) invariant quantities,
defined as the ratios of
magnetic correlation lengths over the lattice size. The correlation length is in fact
not uniquely defined
on a finite lattice. This ambiguity allows for optimization so
as to reduce corrections to scaling.  Using this strategy, we can unbiasedly
find the value of the critical coupling $U_c$ and obtain critical exponents.  The exponents we find for both
models are consistent with the one-loop $\varepsilon$-expansion
\cite{Herbut09a}. Most notably, the anomalous bosonic dimension $\eta$ is
large.  Our results are based on auxiliary-field quantum Monte Carlo (QMC)
simulations on lattices with up to $18 \times 18 $ unit cells.  Since these
lattices sizes are {\it small}, we verify our approach for
the Mott transition of the Kane-Mele-Hubbard model \cite{Rachel10}, which is known to be in
the UC of the three-dimensional (3D) XY model \cite{Hohenadler10,*Hohenadler10_erratum,Hohenadler12,PhysRevLett.107.166806}.

The organization of the paper is the following.  In Sec.~\ref{sec:model}, we
define the models. In Sec.~\ref{sec:fss}, we discuss the finite-size
scaling, and in Sec.~\ref{sec:method} we provide some details about the
QMC method. Section~\ref{sec:results} contains our
results, and Sec.~\ref{sec:summary} provides a summary and the conclusions.
Appendix~\ref{sec:xi} gives details about the definition of a correlation
length in finite systems.
Appendix~\ref{sec:fsschicl} contains an additional finite-size scaling analysis of the Hubbard model on the honeycomb lattice which corroborates the main findings.

\section{Models} \label{sec:model}
\begin{figure}[b]
\includegraphics[width=0.55\linewidth,keepaspectratio]{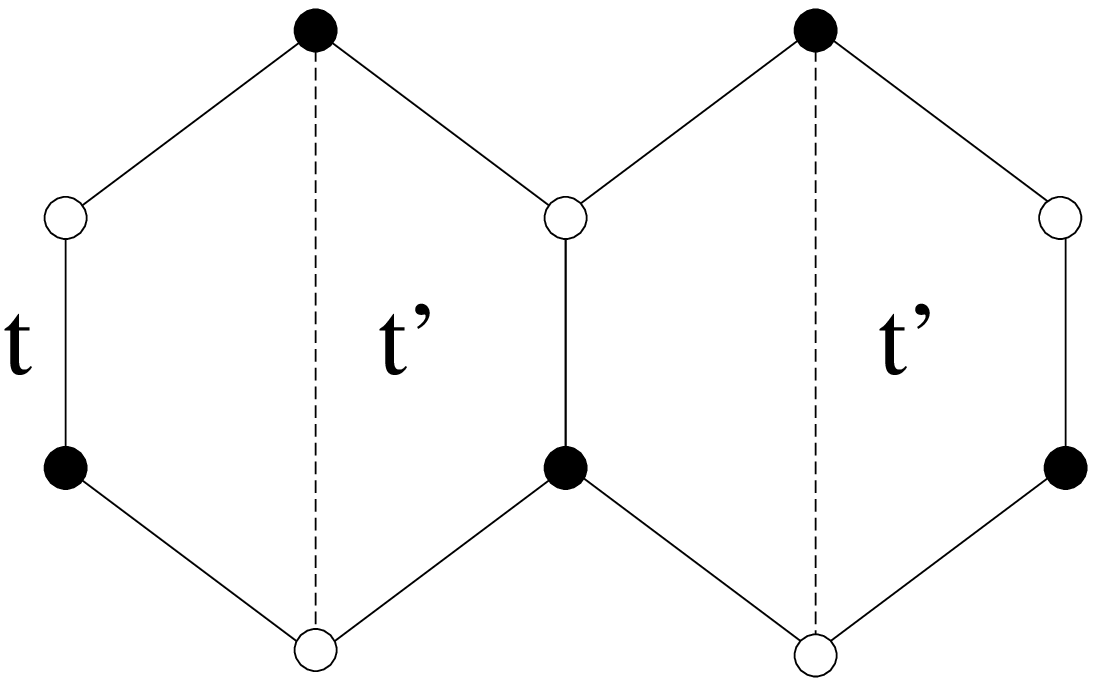}
\caption{Illustration of the hopping term in Eq.~(\ref{kmh}).
  Solid lines represent a nearest-neighbor hopping with amplitude $-t$,
  while dashed lines represent hopping across the hexagon with
  amplitude $-t'$. In this work we consider the cases $t'=0$ (honeycomb
  lattice) and $t'=-t$ ($\pi$-flux lattice). }
\label{hexagonpiflux}
\end{figure}

In this work, we study three different models with a Hubbard repulsion, namely,
the Hubbard model on the honeycomb lattice ({\it honeycomb Hubbard model}),
the Hubbard model on the $\pi$-flux lattice ({\it $\pi$-flux Hubbard model}),
and the Hubbard model on the honeycomb lattice with spin-orbit coupling ({\it
Kane-Mele-Hubbard model}). These models are subsumed by the Hamiltonian
\begin{equation}\label{kmh}
\begin{split}
  {\cal H}
  = 
  &\sum_{ \vec{\imath},\vec{\jmath},\sigma} \hat{c}^{\dagger}_{\vec{\imath},\sigma} T^{\phantom{\dagger}}_{\vec{\imath},\vec{\jmath} }\,\hat{c}^\nag_{\vec{\jmath},\sigma}
  + 
  \rmi\,\lso \sum_{\llas \vec{\imath},\vec{\jmath}\rras}
  \hat{c}^{\dagger}_{\vec{\imath}}\,
  (\vec{\nu}^{\phantom{\dagger}}_{\vec{\imath},\vec{\jmath}} \cdot \vec{\sigma})\,
  \hat{c}^\nag_{\vec{\jmath}}  \, \\
 &+ U\sum_{\vec{\imath}}\left(n_{\vec{\imath},\uparrow}-\frac{1}{2}\right)\left(n_{\vec{\imath},\downarrow}-\frac{1}{2}\right),
\end{split}
\end{equation}
where $\hat{c}^{\dagger}_{\vec{\imath},\sigma}$ is the creation operator for an
electron with spin $\sigma$ at site $\vec{\imath}$ and
$n_{\vec{\imath},\sigma}\equiv\hat{c}^{\dagger}_{\vec{\imath},\sigma}\hat{c}^\nag_{\vec{\imath},\sigma}$
is the corresponding number operator. The first term in Eq.~(\ref{kmh})
corresponds to single-particle hopping between nearest neighbors with
amplitude $-t$, and across hexagons with amplitude $-t'$ (see Fig.~\ref{hexagonpiflux}).  The second term couples next-to-nearest-neighbor sites and represents the intrinsic spin-orbit interaction of
amplitude  $\lambda$. For a hopping process between sites $ \vec{\imath}$ and $\vec{\jmath}$  via site $\vec{r}$,
$\vec{\nu}_{\vec{\imath},\vec{\jmath}}  =   (\vec{r} - \vec{\imath}) \times
  (\vec{\jmath} - \vec{r}) / |  (\vec{r} - \vec{\imath}) \times  (\vec{\jmath} -
    \vec{r}) |  = \pm \vec{e}_z $. 
The spin-orbit term opens a mass gap and leads to a topological band structure
\cite{KaneMele05}. If the $z$ component of spin is conserved, the Kane-Mele model corresponds to two copies of the
Haldane model \cite{Haldane98} with opposite Chern numbers for the up and down spin sectors.
The parameter $U>0$ characterizes the Hubbard on-site
repulsion. We consider the model at zero chemical
potential, corresponding to half-filling. 

If $\lambda=0$ and $t'=0$, Eq.~(\ref{kmh}) becomes the Hamiltonian of the
honeycomb Hubbard model. For $\lambda=0$ and $t'=-t$, it corresponds to the
$\pi$-flux Hubbard model.
The $\pi$-flux lattice emerges in the large-$N$ limit of the Heisenberg-Hubbard model \cite{Affleck88,Assaad04}.
Finally, for $\lambda>0$ and $t'=0$,
Eq.~(\ref{kmh}) corresponds to the Kane-Mele-Hubbard model.

The honeycomb and $\pi$-flux Hubbard models both have a semimetallic ground
state in the noninteracting case. In contrast, the spin-orbit term of the Kane-Mele-Hubbard
model opens a topological band gap even for $U=0$. 

\subsection{Honeycomb and $\pi$-flux Hubbard models ($\lambda=0$)}

For $\lambda=0$  and $t'/t = 0,-1$, the first term in Eq.~(\ref{kmh})
gives rise to a band structure of massless Dirac fermions.  At $t'=0$,  the
two inequivalent cones are located at the Brillouin zone boundaries.  As  a
function of $t'/t$,  the cones meander (since the $C_3$ symmetry is
broken), and are located at
\begin{equation}
  \vec{K} =  \pm 4 \,{\rm arccos} \left( -\frac{(1 + t'/t)}{2}  \right) (\vec{b}_1 + \vec{b}_2/2)\,,
\end{equation}
where $\vec{b}_1=(1,-1/\sqrt{3})$ and $\vec{b}_2=(0,2/\sqrt{3})$.
For the values of $t'$ considered here, the 
cones are pinned to specific $\vec{K}$ points due to lattice symmetries. For
$t'=0$, we have the $C_3$ symmetry of the honeycomb lattice, whereas for $t'/t=-1$
we have the $C_4$ symmetry of the $\pi$-flux lattice. 
Expanding around $\vec{K}$ gives the spectrum
\begin{equation}
    E(\vec{K}  + \vec{k} )   = \pm \sqrt{ (v_x k_x )^2     + (v_y k_y )^2  } +O(k)^2, \quad \vec{k}\rightarrow 0
\end{equation}
with velocities
\begin{equation}
  v_x  =   t  \sqrt{ 1 - \frac{(1 + t'/t)^2}{4} }\,, \quad 
  v_y  =   t  \frac{\sqrt{3}\left| 1- t'/t\right|}{2} \,.
\label{velocities}
\end{equation}

\begin{figure}
\includegraphics[width=\linewidth,keepaspectratio]{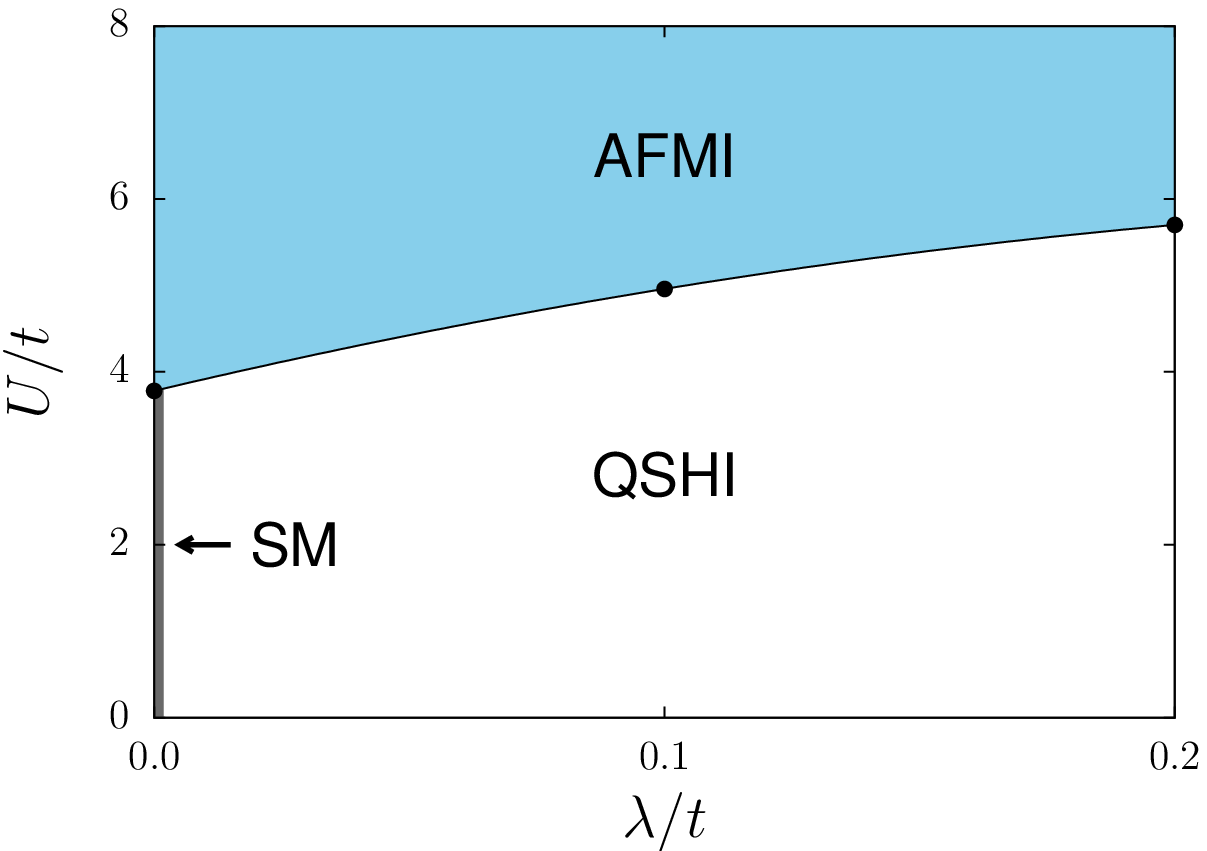}
\caption{(Color online) Phase diagram of the Kane-Mele-Hubbard model ($\lambda>0$) and the honeycomb
  Hubbard model ($\lambda=0$) from QMC simulations, taken from Ref.~\cite{PhysRevB.90.085146}. The
  phases correspond to a semimetal (SM), an antiferromagnetic Mott insulator
  (AFMI), and a quantum spin-Hall insulator (QSHI).}
\label{phase}
\end{figure}

At $T=0$, both the honeycomb and the $\pi$-flux Hubbard models
are believed to describe a continuous phase transition between a semimetallic
phase that is adiabatically connected to $U=0$, and an insulating
antiferromagnetic phase at large values of $U$. This phase transition has
prompted numerous studies, in particular concerning the possible presence of
an intermediate spin-liquid phase \cite{Meng10,PhysRevB.83.024408,*PhysRevB.85.045123}. In line with subsequent
studies \cite{Sorella12,Assaad13}, we show in the following that the
phase transition is described by the Gross-Neveu-Heisenberg UC
\cite{Herbut06,Herbut09,Herbut09a}. In this scenario, the two phases
are separated by a single critical point without any intermediate phase. For
the honeycomb Hubbard model, the phase diagram from QMC simulations is shown
in Fig.~\ref{phase}, where it corresponds to the $\lambda=0$ axis.

The phase transition is characterized by the $O(3)$ antiferromagnetic order
parameter
\begin{equation}
  \vec{\phi}(\vec{x})= \vec{S}(\vec{x}_A) - \vec{S}(\vec{x}_B)\,,
\label{ophubbard}
\end{equation}
where $\vec{x}$ is a site of a triangular lattice that corresponds
to an elementary unit cell of the honeycomb lattice, and $\vec{x}_A$ and
$\vec{x}_B$ are lattice sites (in the same unit cell) that belong to the
$A$ and $B$ sublattices, respectively.

\subsection{Kane-Mele-Hubbard model ($\lambda\neq0$)}

In Fig.~\ref{phase}, we show the phase diagram of the Kane-Mele-Hubbard model
from QMC simulations \cite{PhysRevB.90.085146}. The model exhibits three
phases, separated by second-order transition lines. For $\lambda=0$, the
model reduces to the honeycomb Hubbard model, see above. A nonzero $\lambda$
opens a gap at the Dirac points, and leads to the formation of a quantum spin
Hall insulator \cite{KaneMele05,KaneMele05b}. At large $U$, the model
describes an antiferromagnetic Mott insulator with magnetic order in the
transverse spin direction \cite{Rachel10,Hohenadler10,Zheng11,Hohenadler12}.
The Kane-Mele-Hubbard model has been studied in great detail to understand
correlation effects in topological insulators \cite{HoAsreview2013}.

The spin-orbit interaction reduces the symmetry of the Kane-Mele-Hubbard
model to the $O(2)$ group. Consequently, the quantum  phase transition between the
quantum spin Hall phase and the antiferromagnetic Mott insulator
belongs to the well known 3D XY UC
\cite{Hohenadler12,PhysRevLett.107.166806}. It is characterized by
the $O(2)$ antiferromagnetic order parameter
\begin{equation}
  \vec{\phi}( \vec{x})= (S^x(\vec{x}_A), S^y(\vec{x}_A)) - (S^x(\vec{x}_B),
  S^y(\vec{x}_B)) \,.
\label{opkm}
\end{equation}
In the following, we set $t=1$.

\section{Finite-Size Scaling}\label{sec:fss}

Finite-size-scaling (FSS) theory is a powerful method that allows one to study
the critical behavior of models using finite-size data. Unlike
infinite-volume methods, FSS is concerned with analyzing the scaling behavior in a
regime where the correlation length $\xi$ and the linear size of the system
$L$ are of comparable size, $\xi\sim L$ \cite{barber83,cardyfss,privmanfss,PV-02}. To be precise, FSS
theory allows one to formulate the scaling behavior of the observables in the
so-called FSS limit, where $L,\xi\rightarrow\infty$, at fixed $\xi/L$. The
FSS method has been recently discussed in the context of quantum phase
transitions in Ref.~\cite{CPV-14}. 

We consider the spatial two-point correlation function $\C(\vec{x}-\vec{y})$
of the order parameter $\phi(\vec{x})$ at $T=0$,
\begin{equation}
\C(\vec{x}-\vec{y}) \equiv \langle\vec{\phi}(\vec{x})\cdot\vec{\phi}(\vec{y})\rangle.
\end{equation}
Using the spatial correlations $\C(\vec{x})$ one can define various
observables, the FSS behavior of which allows one to study the critical
properties of second-order phase transitions. We study the zero-momentum
Fourier transform of the two-point function $\chicl$, defined as
\begin{equation}
\chicl(U,L)\equiv\sum_{\vec{x}}\C(\vec{x}).
\label{chidef}
\end{equation}
Close to a second-order phase transition at $U=U_c$, $\chicl$
exhibits the following FSS behavior \cite{CPV-14}
\begin{equation}
\label{chifss}
\chicl(U,L) = L^{2-z-\eta}\left[f_\chicl(w)+L^{-\omega}g_\chi(w)\right] + B(U)\,,
\end{equation}
\begin{equation}
\label{wdef}
w\equiv uL^{1/\nu}, \qquad u\equiv ({U-U_c})/U_c\,,
\end{equation}
where $\nu$, $\eta$, and $z$ are universal critical exponents, $\omega$ is a
generic correction-to-scaling exponent and $B(U)$ is a nonuniversal analytic
background term that originates from the nonuniversal, short-distance
behavior of $\C(x)$, i.e., from the terms in the sum of Eq.~(\ref{chidef})
for which $|\vec{x}|\ll L$.
According to RG theory, corrections to scaling may have several origins (see
also Ref.~\cite{CPV-14}):

(i) Irrelevant operators give rise to scaling corrections with an exponent
  $\omega$ equal to their negative RG dimension.

(ii) Analytical scaling corrections originate from the so-called nonlinear
  scaling fields \cite{PhysRevB.27.4394}, according to which the scaling
  fields are replaced by a generic analytical expansion in the Hamiltonian
  parameters. For instance, $u$ in Eq.~(\ref{wdef}) should be replaced by
  an expansion of the form $u+cu^2+o(u^2)$, where $c$ is a nonuniversal
  constant, resulting in a scaling correction with exponent $\omega=1/\nu$.

(iii) Additional scaling corrections arise from the analytic part of the
  free energy. This is the case of the background term $B(U)$, which can be
  considered as a subleading term with an effective correction-to-scaling
  exponent $\omega=2-z-\eta$.

In general, one expects several correction-to-scaling terms, the leading one
being the one with the smallest exponent $\omega$. Here and in the following,
we consider the leading scaling correction only.

RG-invariant quantities (also called phenomenological couplings) are instrumental for investigating the critical behavior. Here, we consider ratios of the correlation length and the lattice size $L$. As explained in Appendix~\ref{sec:xi}, on a finite lattice there is no unique definition of the correlation length. We defined several correlation lengths that mimic the definition of the second-moment correlation length of the two-point function $\C(\vec{x})$; all these quantitites are observables that scale as $\propto L$ in the FSS limit, so that their ratio with the lattice size $L$ is RG-invariant. We consider
\begin{eqnarray}
  \label{Rxi1-def}
  \Rxii{1}(U,L) &\equiv& \xi^{(1)}(U,L)/L\,,\\
  \label{Rxi2-def}
  \Rxii{2}(U,L) &\equiv& \xi^{(2)}(U,L)/L\,,\\
  \label{Rxi-def}
  \Rxi(U,L) &\equiv& \xi(U,L)/L\,,\\
  \label{Rxis-def}
  \Rxis{\kappa}{\rho}(U,L) &\equiv& \xi_{s,\kappa,\rho}(U,L)/L\,,
\end{eqnarray}
where $\xi^{(1)}$, $\xi^{(2)}$ are two finite-size correlation lengths
defined in terms of the Fourier transform of $\C(\vec{x})$ and corresponding
to the two principal directions, $\xi$ is a generalized $f$-mean value of
$\xi^{(1)}(L)$ and $\xi^{(2)}(L)$, and $\xi_{s,\kappa,\rho}$ is a correlation
length defined in terms of the two-point function $\C(\vec{x})$ in real
space. These correlation lengths are inequivalent observables in the FSS
limit; their definitions are discussed in Appendix~\ref{sec:xi}. The
parameters $\kappa$ and $\rho$ that enter in the definition of
$\xi_{s,\kappa,\rho}$ are scale-invariant ratios that influence the amplitude
of the scaling corrections (see Appendix~\ref{sec:xi:real}). 

As discussed in
Sec.~\ref{sec:method}, our simulation data for the $\pi$-flux Hubbard model are for
lattices with  $L_1=L/2$ unit cells in direction $1$ and $L_2=L$
unit cells in direction $2$. In view of the anisotropy of the lattice, we 
use a slightly different definition for the RG-invariant quantity $\Rxii{1}$:
\begin{equation}
\label{Rxispi-def}
\Rxii{1}(U,L)\equiv \xi^{(1)}(U,L)/(L/2) \quad\text{($\pi$-flux lattice)}.
\end{equation}

According to FSS theory, a generic RG-invariant observable $R(U,L)$ obeys the scaling ansatz
\begin{equation}
R(U,L) = f_R(w) + L^{-\omega}g_R(w),
\label{RG-scaling}
\end{equation}
where the function $f_R(w)$ is universal, apart from a nonuniversal
normalization of the scaling variable $w$. Aside from depending on the
UC of the phase transition, $f_R(w)$ also depends on the
boundary conditions of the system and on the aspect ratio. In
Eq.~(\ref{RG-scaling}), we have included a correction-to-scaling term
$L^{-\omega}g_R(w)$, which decays with a correction-to-scaling exponent
$\omega$.

As illustrated in Appendix~\ref{sec:xi}, the finite-size correlation lengths
$\xi^{(1)}$, $\xi^{(2)}$, $\xi$, and $\xi_{s,\kappa,\rho}$ are computed with a ratio that
involves $\chicl$ [see Eq.~(\ref{chidef})]. Therefore, scaling corrections for $\Rxii{1}(U,L)$,
$\Rxii{2}(U,L)$, $R_\xi$, and $\Rxis{\kappa}{\rho}$ are analogous to those of
$\chicl$. In particular, they are also affected by scaling corrections that decay
with an exponent $\omega=2-z-\eta$ and originate from the analytic part of
the free energy.

A popular method for extracting the critical coupling $U_c$ from the FSS
behavior of a model is the so-called crossing method. It is based on the
observation that, neglecting scaling corrections in Eq.~(\ref{RG-scaling})
(i.e., taking $\omega\rightarrow\infty$), the equation
\begin{equation}
  R(U,L)=R(U,L')
  \label{crossdef_gen}
\end{equation}
admits a solution for $U=U_c$, i.e., $u=0$. If in an interval around $u=0$
the scaling function $f_R(w)$ is monotonic, then, locally, this is the only
solution to Eq.~(\ref{crossdef_gen}). This implies that the curves $R(U,L)$
as a function of $U$ intersect at $U=U_c$ for all lattice sizes
$L$. Typically, one observes instead a drift in the crossings, which is due to
scaling corrections. To determine the critical coupling $U_c$, one usually
defines a pseudocritical coupling $U_{c,R}(L)$ as the solution of
Eq.~(\ref{crossdef_gen}) with $L'=\alpha L$, where $\alpha$ is a fixed
ratio. Here, the available lattice sizes do not allow us to use this
definition for $U_{c,R}(L)$. Instead, we define a pseudocritical coupling
$U_{c,R}(L)$ as the solution of Eq.~(\ref{crossdef_gen}) with $L'=L+c$, that is,
\begin{equation}
R(U_{c,R}(L),L)=R(U_{c,R}(L),L+c),
\label{crossdef}
\end{equation}
where $c$ is a fixed constant. By inserting Eq.~(\ref{RG-scaling}) in
Eq.~(\ref{crossdef}), and expanding for $L\rightarrow\infty$, one can show
that for  $L\rightarrow\infty$ $U_{c,R}(L)\rightarrow U_c$ according to
\begin{equation}
U_{c,R}(L) = U_c +A L^{-e}, \qquad e=1/\nu+\omega,
\label{pseudoconv}
\end{equation}
where $A$ is a nonuniversal constant.  Using different RG-invariant
quantities, we can define different pseudocritical couplings $U_{c,R}(L)$ that
all converge to $U_c$ for $L\rightarrow\infty$. This property can be
used to corroborate the result for $U_c$.

\section{Quantum Monte Carlo method}
\label{sec:method}

We used the projective auxiliary-field QMC algorithm to compute the
spin-spin correlations. Because a detailed discussion of the algorithm is
beyond the scope of this work,  we refer the reader to
 Refs.~\cite{AssaadEvertz_rev,Hohenadler12}.  

Ground-state expectation values
of observables are calculated according to the equation
\begin{equation}
\langle \hat{O} \rangle_0   = \lim_{\Theta \rightarrow \infty} \frac{\langle \Psi_T | e^{-\Theta \hat{H}} \hat{O} e^{-\Theta \hat{H}} |  \Psi_T \rangle } 
   { \langle \Psi_T | e^{-2\Theta \hat{H}}  |  \Psi_T \rangle   }\,,
\end{equation}
where the ground-state wave function is filtered out of a trial wave
function (required to be nonorthogonal to the ground state) by projection
along the imaginary-time axis.  The QMC algorithm relies on a Trotter
decomposition. We used a symmetric version that produces a
systematic error of the order $(\Delta \tau)^2 $, where $\Delta \tau $
is the imaginary-time step.  We typically used $\Delta \tau= 0.1$, and a projection parameter $\Theta = 30$. The trial wave function
was taken to be the ground state of the noninteracting Hamiltonian and
chosen to be a spin singlet.  The method has two sources of systematic
errors: the projection parameter and the high-energy (or short
imaginary-time) cutoff $\Delta \tau$.  For a given statistical precision of
$0.1\%$ for the antiferromagnetic order parameter, we checked that the
choice of the projection parameter and trial wave function guarantees
convergence to the ground state.  On the other hand, at $U_c = 3.8$ and for
the honeycomb lattice, the finite value of $\Delta \tau $ leads to a
systematic error of the order of $0.5\%$. This high-energy cutoff may
slightly shift the critical values of $U$ at which the transition occurs
but should not alter the universality.  Finally, we used an SU(2)-symmetric
Hubbard-Stratonovich transformation \cite{AssaadEvertz_rev} to ensure that
this symmetry is conserved for each field configuration.

For the simulations on the honeycomb lattice we used lattices spanned by the vectors $ \vec{L}_1= L \vec{a}_1 $
and $ \vec{L}_2 = L \vec{a}_2 $,
where $\vec{a}_1=(1,0)$ and $\vec{a}_2=(1/2,\sqrt{3}/2)$,
and with boundary conditions $ c_{\vec{\imath} +
  \vec{L}_n, \sigma} = c_{\vec{\imath}, \sigma} $ with $n = 1,2$.  With this
choice of boundary conditions, and the values of $L$ as multiples of $3$, the Dirac points are part of the reciprocal lattice.

\begin{figure}
\includegraphics[width=\linewidth,keepaspectratio]{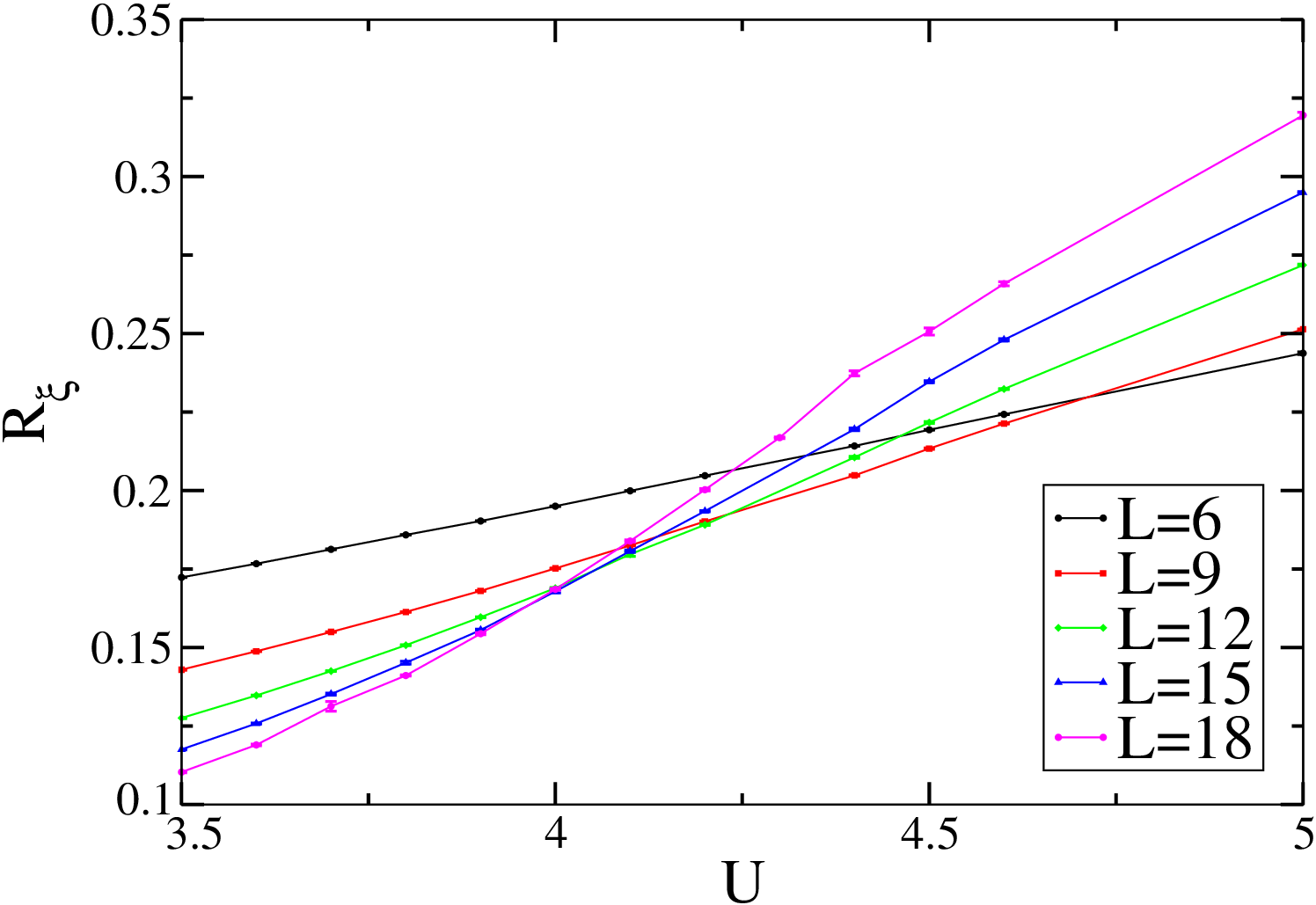}
\caption{(Color online) RG-invariant quantity $\Rxi$ for the honeycomb
  Hubbard model. Lines are guides to the eye.}
\label{xil_hubbard}
\end{figure}

\begin{figure}[b]
\vspace{1em}
\includegraphics[width=\linewidth,keepaspectratio]{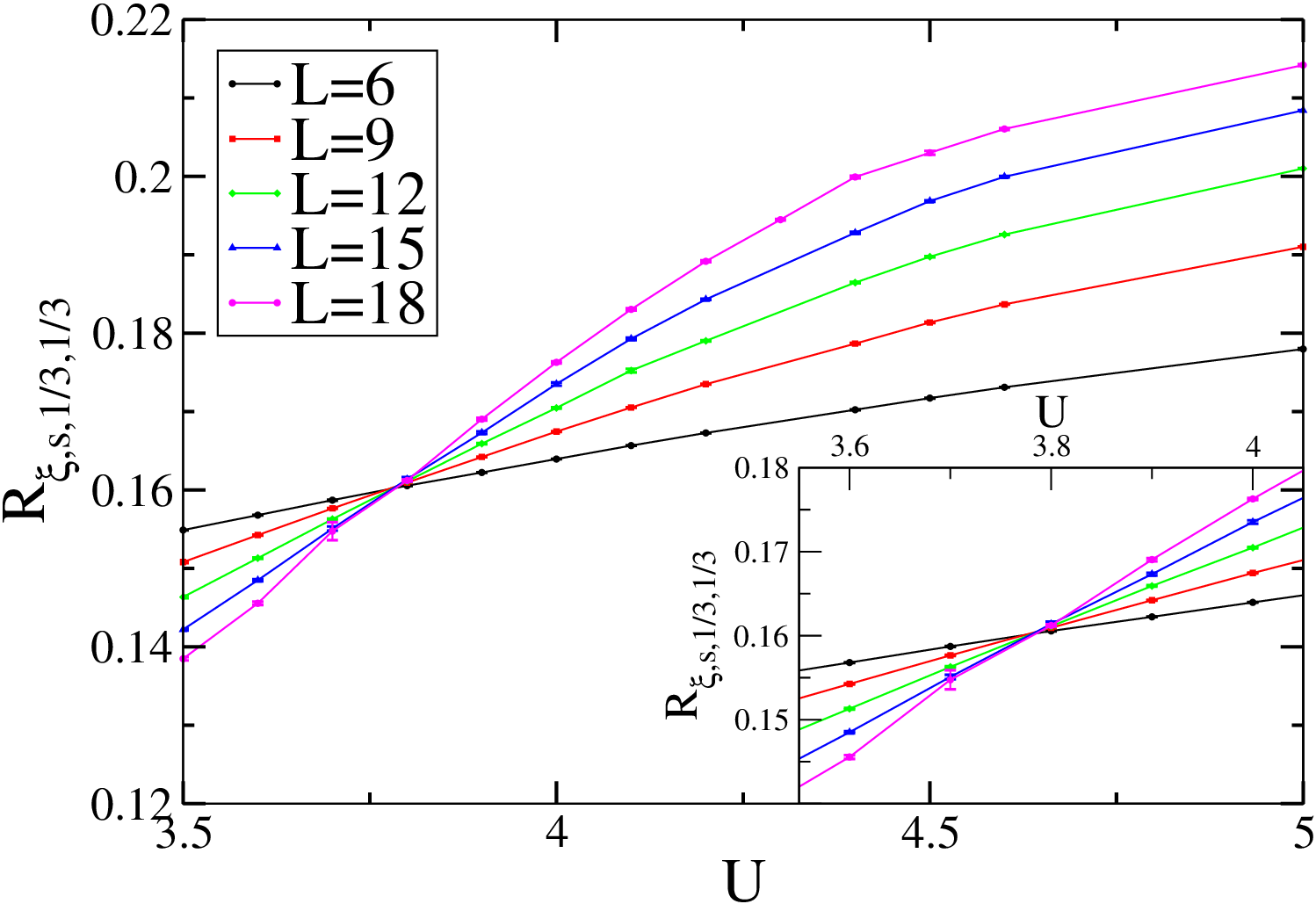}
\caption{(Color online)  Same as Fig.~\ref{xil_hubbard} for $\Rxist$. Inset: magnification of the data close to their crossing at $U\approx 3.8$.}
\label{sumxil_hubbard}
\end{figure}

For the $\pi$-flux lattice we considered lattices defined by the vectors $
\vec{L}_1 = \frac{L}{2} \vec{a}_1 $ and $ \vec{L}_2 = \frac{L}{2} ( 2
\vec{a}_2 - \vec{a}_1 ) $, again with boundary conditions $ c_{\vec{\imath} +
  \vec{L}_n, \sigma} = c_{\vec{\imath}, \sigma} $ .  This choice of boundary
conditions is equivalent to a lattice that extends over $L_1=L/2$ unit
cells in the $\vec{a}_1$ direction and over $L_2=L$ unit cells in the
$\vec{a}_2$ direction. The total number of two-site unit cells is $L\times
L/2$, and the total number of lattice sites is $L\times L/2 \times 2=L\times
L$.  This also makes the lattice equivalent to an $L\times L$ square lattice.
For $L$ being a multiple of $4$ the Dirac points are part
of the reciprocal lattice.

\section{Results}\label{sec:results}

\subsection{Honeycomb Hubbard model}\label{sec:results:hubbard}

We simulated the honeycomb Hubbard model on lattices with
$L=6$, $9$, $12$, $15$, and $18$. As discussed in Appendix~\ref{sec:xi:real}, the
correlation length $\xi_{s,\kappa,\rho}$ is computed for $\kappa=\rho=1/3$
only.  In Figs.~\ref{xil_hubbard} and \ref{sumxil_hubbard} we show the
RG-invariant quantities $\Rxi(U,L)$ and $\Rxist(U,L)$ as a
function of $U$ and for lattice sizes $L=6-18$. We observe that the curves of
$\Rxi(U,L)$ for different $L$ do not show a common intersection point, but
exhibit a systematic drift of the intersection points from $U\approx 4.7$
(the crossing point of the curves for $L=6$ and $L=9$) towards smaller values
of $U$; the data for $\Rxi(U,L)$ and for the two largest lattice sizes
intersect at $U\approx 3.9-4$. The curves of $\Rxist(U,L)$ shown in
Fig.~\ref{sumxil_hubbard} exhibit instead a common intersection at $U\approx
3.8$.

\begin{figure}
\includegraphics[width=\linewidth,keepaspectratio]{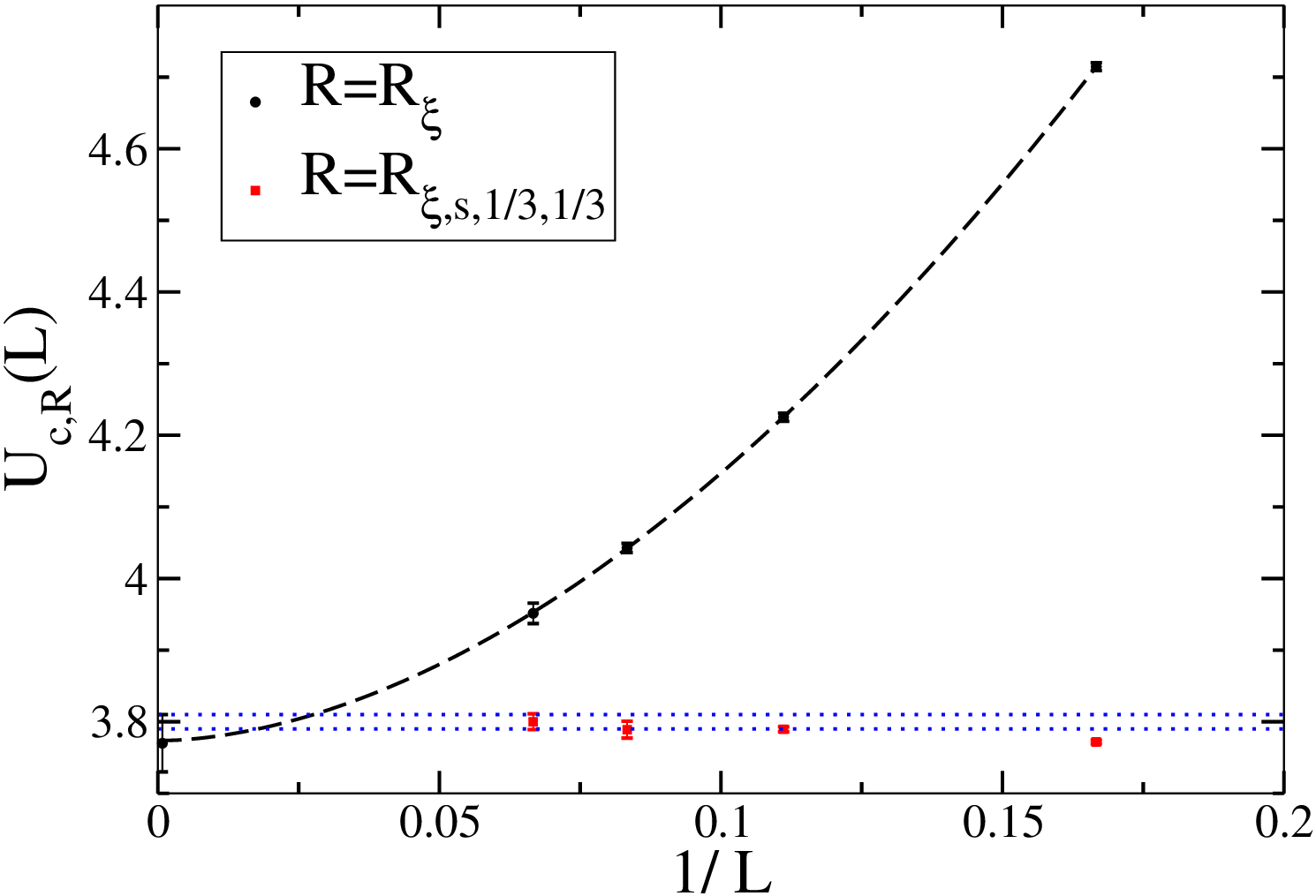}
\caption{(Color online) pseudocritical coupling $U_{c,R}$ for the honeycomb Hubbard
  model, obtained by numerically solving Eq.~(\ref{crossdef}) for two
  phenomenological couplings $R=\Rxi$ and $R=\Rxist$. The plotted value of
  $U_{c,\Rxi}=3.77(4)$ for $L\rightarrow\infty$ has been obtained by fitting
  the data to Eq.~(\ref{pseudoconv}). The dashed line represents the
  right-hand side of Eq.~(\ref{pseudoconv}), with central values of the fit
  $U_c=3.77$, and $e=1.8$. The dotted lines indicate the interval in the
  final estimate of the critical coupling $U=3.80(1)$ as reported in
  Eq.~(\ref{Uc_hubbard}).}
\label{cross_hubbard}
\end{figure}

These observations are confirmed by the analysis of the pseudocritical
coupling $U_{c,R}(L)$. In Fig.~\ref{cross_hubbard}, we show $U_{c,R}(L)$ as a
function of $1/L$, as obtained by numerically solving Eq.~(\ref{crossdef}),
with $R=\Rxi$, $\Rxist$ and $c=3$. For each pair of lattice sizes $L$ and
$L+3$ we fitted the data for $R_\xi$ and $\Rxist$ to a suitable Taylor expansion in
$U$ in an interval around the crossing point. These fits provide an
interpolation of the curves for $R(U,L)$ and $R(U,L+3)$ that, in turn, allows
us to solve Eq.~(\ref{crossdef}). The resulting error bar of $U_R(L)$, which
is determined from the covariance matrix of the coefficients of the fits used
to interpolate $R(U,L)$, may underestimate the uncertainty in $U_R(L)$
because it does not take into account a possible systematic error in the
truncation of the Taylor expansion of $R(U,L)$.  Figure~\ref{cross_hubbard}
reveals that $U_{c,\Rxi}(L)$ decreases slowly upon increasing $L$, whereas
$U_{c,\Rxist}(L)$ remains stable; for $L\ge 9$, $U_{c,\Rxist}(L)$ is constant
within error bars.  In order to
extrapolate $U_c$ from the pseudocritical coupling $U_{c,R_\xi}(L)$, we 
fitted the data for $U_{c,R_\xi}(L)$ to the right-hand side of Eq.~(\ref{pseudoconv}),
leaving $U_c$, $A$, and the exponent $e$ as free parameters.  The fitted
values are $U_c=3.77(4)$ and $e=1.8(1)$, with $\chi^2/\DOF=0.02$ (DOF: degrees
of freedom). Within the statistical precision, the result for $U_c=3.77(4)$
is in full agreement with the pseudocritical couplings $U_{c,\Rxist}(L)$ for
all available lattice sizes.  In Fig.~\ref{cross_hubbard}, we also show the
right-hand side of Eq.~(\ref{pseudoconv}) (the dashed line), which illustrates the convergence of
$U_{c,R_\xi}(L)$ to the critical coupling $U_c$ for $L\rightarrow\infty$.

\begin{table*}
\caption{Results of the fits of $R=\Rxist$ for the honeycomb Hubbard model to
  Eq.~(\ref{RG-scaling-exp}) (first three sets) and to
  Eq.~(\ref{RG-scaling-exp-omega}) (last three sets), with $U\in
  [3.6,4]$. $L_{\rm min}$ is the minimum lattice size taken into account in
  the fits.}
\begin{ruledtabular}
\begin{tabular}{r@{\hspace{1em}}l@{\hspace{2em}}.@{\hspace{4em}}.@{\hspace{4em}}.@{\hspace{4em}}.}
& \multicolumn{1}{c}{$L_{\rm min}$} & \multicolumn{1}{c}{$U_c$} & \multicolumn{1}{c}{$\nu$} & \multicolumn{1}{c}{$\Rxist^*$} & \multicolumn{1}{c}{$\chi^2/\DOF$} \\
\hline
\multirow{4}{*}{$n_{\rm max}=1$} & $6$  & 3.782(1)   & 0.758(4) & 0.16017(3) & 443.2/21 \\
 & $9$  & 3.7954(15) & 0.816(7) & 0.16077(6) & 39.5/16 \\
 & $12$ & 3.7975(30) & 0.87(2)  & 0.1609(2)  & 17.8/11 \\
 & $15$ & 3.798(9)   & 0.91(5)  & 0.1610(6)  & 9.5/6 \\
\\
\multirow{4}{*}{$n_{\rm max}=2$} & $6$  & 3.775(1)   & 0.747(4) & 0.16004(3) & 331.0/20 \\
& $9$  & 3.790(2)   & 0.812(7) & 0.16063(7) & 18.0/15 \\
& $12$ & 3.792(3)   & 0.86(2)  & 0.1607(2)  & 5.0/10 \\
& $15$ & 3.797(8)   & 0.87(5)  & 0.1610(6)  & 3.4/5 \\
\\
\multirow{4}{*}{$n_{\rm max}=3$} & $6$  & 3.780(1)   & 0.694(6) & 0.16014(3) & 240.0/19 \\
& $9$  & 3.791(2)   & 0.786(15)& 0.16066(7) & 14.7/14 \\
& $12$ & 3.792(4)   & 0.85(3)  & 0.1607(2)  & 4.9/9 \\
& $15$ & 3.797(8)   & 0.86(6)  & 0.1610(6)  & 3.3/4 \\
\\
$n_{\rm max}=2$ & $6$  &  3.823(4)  & 0.755(4) &  0.175(1) &  167.4/19 \\
$m_{\rm max}=0$ & $9$  &  3.805(11) & 0.813(7) &  0.167(5) &  16.0/14 \\
$\omega=0.15$   & $12$ &  3.82(5)   &  0.86(2) &  0.18(3)  &   4.6/9 \\
\\
$n_{\rm max}=2$ & $6$  &  3.820(4)  & 0.754(4) &  0.1679(6) & 166.5/19 \\
$m_{\rm max}=0$ & $9$  &  3.804(10) & 0.813(7) &  0.164(2)  & 16.01/14 \\
$\omega=0.3$ & $12$ &  3.82(4)   & 0.86(2)  &  0.168(14) &  4.6/9 \\
\\
$n_{\rm max}=2$ & $6$  &  3.816(3) &  0.754(4) &  0.1653(4)  & 165.7/19 \\
$m_{\rm max}=0$ & $9$  &  3.803(9) &  0.813(7) &  0.1629(16) & 16.0/14 \\
$\omega=0.45$ & $12$ &  3.82(4)  &  0.86(2)  &  0.166(9) &   4.6/9
\end{tabular}
\end{ruledtabular}
\label{fit_sumxil_hubbard}
\end{table*}

The slow convergence of $U_{c,R_\xi}(L)$ to $U_c$ implies that $R_\xi$ is
affected by large scaling corrections. As discussed in Sec.~\ref{sec:fss},
these can stem from various sources. As shown in the
analysis below, the critical behavior belongs to the Gross-Neveu-Heisenberg
UC. Using functional RG methods, the leading irrelevant operator in this UC
has been determined as $\omega\approx 0.9$ \cite{PhysRevB.89.205403}. In the present model, an additional irrelevant operator is associated
with the restoration of the Lorentz symmetry; within the
$\varepsilon$-expansion, its negative dimension is
$\omega=\frac{4}{5}\varepsilon$ \cite{Herbut09a}, where one should set
$\varepsilon=1$ for the two-dimensional system considered here. Although such a simple
substitution has to be taken with some care, we have no reason
to presume the existence of an irrelevant operator with a small $\omega$
exponent. Analytical scaling corrections arising from nonlinear scaling
fields are also not expected to play an important role here. Indeed, as we
show in the following, $\nu\lesssim 1$, so that scaling corrections $\propto L^{-1/\nu}$
are not particularly large.  On the other hand, in the Gross-Neveu picture,
the dynamical exponent $z$ is equal to 1, and field theoretical methods indicate a
large $\eta$ exponent. Within the first-order $\varepsilon$-expansion, one
has $\eta=\frac{4}{5}\varepsilon$ \cite{Herbut09a}, so that by setting
$\varepsilon=1$ one obtains a rather large value of the $\eta$ exponent,
$\eta=0.8$. A large $\eta$ exponent is also confirmed by the analysis
below. Therefore, we expect the zero-momentum Fourier transform of the
two-point function $\chicl$ as well as the RG-invariant quantities $R_\xi$,
$R_{\xi,s}$ to be affected by slowly-decaying scaling corrections, with
$\omega=2-z-\eta\approx 0.2$. However, the amplitude of such scaling
corrections is not universal and also depends on the specific observable.
The stable crossing point observed in Fig.~\ref{sumxil_hubbard} indicates
that the correction to scaling $\propto L^{-0.2}$ is in fact suppressed in
$\Rxist$, i.e., $\Rxist$ is effectively an (approximately) ``improved''
observable \footnote{We notice that the construction of improved observables,
  as well as improved models, where leading scaling corrections are
  suppressed requires in general a fine-tuning of an irrelevant parameter,
  see, e.g., the discussion in Ref.~\cite{HPPV-07}.}.

In view of these results, we determined the critical exponent $\nu$ and
the critical coupling $U_c$ by exploiting the FSS behavior of
$\Rxist$. Following a procedure analogous to the one employed in
Ref.~\cite{PTPV-09}, we fitted $\Rxist$ to a Taylor expansion of
Eq.~(\ref{RG-scaling}). We restricted the analysis to the data where $U$
belongs to an interval $[3.6,4]$ centered at $U=3.8$, which is the approximate common
intersection of the curves in Fig.~\ref{sumxil_hubbard}.  Within this
interval we can expand the scaling function $f_R(w)$ for $R=\Rxist$ in powers
of $w$. Using Eq.~(\ref{wdef}) in Eq.~(\ref{RG-scaling}) and neglecting
scaling corrections, we obtain
\begin{equation}
R = R^* + \sum_{n=1}^{n_{\rm max}}a_n(U-U_c)^nL^{n/\nu}.
\label{RG-scaling-exp}
\end{equation}
We fitted the data for $R=\Rxist$ to Eq.~(\ref{RG-scaling-exp}), leaving
the universal critical value $\Rxist^*\equiv R^*$, the coefficients
$\{a_i\}$, $U_c$, and $\nu$ as free parameters. In order to monitor the role
of the neglected scaling corrections, we repeated the fits disregarding
systematically the smallest lattice sizes. Moreover, to check the reliability
of the Taylor expansion in Eq.~(\ref{RG-scaling-exp}), we repeated the
fit for $n_{\rm max}=1$, $2$, and $3$.

In Table~\ref{fit_sumxil_hubbard} we report the fit results as a function of
the minimum lattice size $L_{\rm min}$ taken into account, and the
expansion order $n_{\rm max}$. Table~\ref{fit_sumxil_hubbard} reveals that
$\chi^2/\DOF$ decreases significantly between
$n_{\rm max}=1$ and $n_{\rm max}=2$, but only marginally
between $n_{\rm max}=2$ and $n_{\rm max}=3$. This indicates that within the
available numerical precision, the range of $U$ considered here
does not allow a linear approximation of $f_{\Rxist}(w)$, whereas a quadratic
approximation appears to be adequate.  Thus, we can restrict the discussion
of the results to the case $n_{\rm max}=2$. The corresponding fits show a
good $\chi^2/\DOF$ for $L_{\rm min}\ge 9$; only for $L_{\rm min}=6$ we have
a large $\chi^2/\DOF$, indicating sizable scaling corrections. Moreover, the
fitted parameters appear to be rather stable upon increasing $L_{\rm min}$. A
conservative judgment of the fit results would give the estimates
$U_c=3.793(5)$, $\nu=0.84(4)$, and $\Rxist^*=0.1608(2)$; these values agree
with the results for $L_{\rm min}=9$, $12$, including a variation of one
error bar, and with the central value of the less precise fit results for
$L_{\rm min}=15$.  As a further check of the reliability of these results, we
repeated the fits with a smaller interval in $U$ where a linear
approximation of $f_{\Rxist}(w)$ is reliable. In
Table~\ref{fit_sumxil_linear_hubbard} we report the results of the fits of
$\Rxist$ to Eq.~(\ref{RG-scaling-exp}) with $n_{\rm max}=1$ and
$U\in [3.7,3.9]$. For $L_{\rm min}\ge 9$, these results display a good
$\chi^2/\DOF$ and are in full agreement with the estimates of $U_c$, $\nu$,
and $\Rxist^*$ given above. These estimates were obtained by an FSS
analysis that neglects scaling corrections. As discussed in the following,
the inclusion of scaling corrections results in slightly
less precise estimates for $U_c$ and $\Rxist^*$.

\begin{table}
\caption{Same as Table~\ref{fit_sumxil_hubbard} for $U\in [3.7,3.9]$ and $n_{\rm max}=1$.}
\begin{tabular}{l@{\hspace{-2em}}d@{\hspace{-1em}}d@{\hspace{-1em}}d@{\hspace{1em}}d}
\hline
\hline
$L_{\rm min}$ & \multicolumn{1}{c}{\hspace{4em}$U_c$} & \multicolumn{1}{c}{\hspace{3em}$\nu$} & \multicolumn{1}{c}{\hspace{3em}$\Rxist^*$} & \multicolumn{1}{c}{$\chi^2/\DOF$} \\
\hline
$6$         & 3.7809(15) & 0.74(1)  & 0.16020(5) & 140.0/11 \\
$9$         & 3.792(2)   & 0.80(2)  & 0.16069(9) & 8.1/8 \\
$12$        & 3.794(5)   & 0.87(5)  & 0.1608(3)  & 2.6/5 \\
$15$        & 3.80(1)    & 0.75(12) & 0.1613(8)  & 1.6/2 \\
\hline
\hline
\end{tabular}
\label{fit_sumxil_linear_hubbard}
\end{table}

\begin{figure}
\vspace{2em}
\includegraphics[width=\linewidth,keepaspectratio]{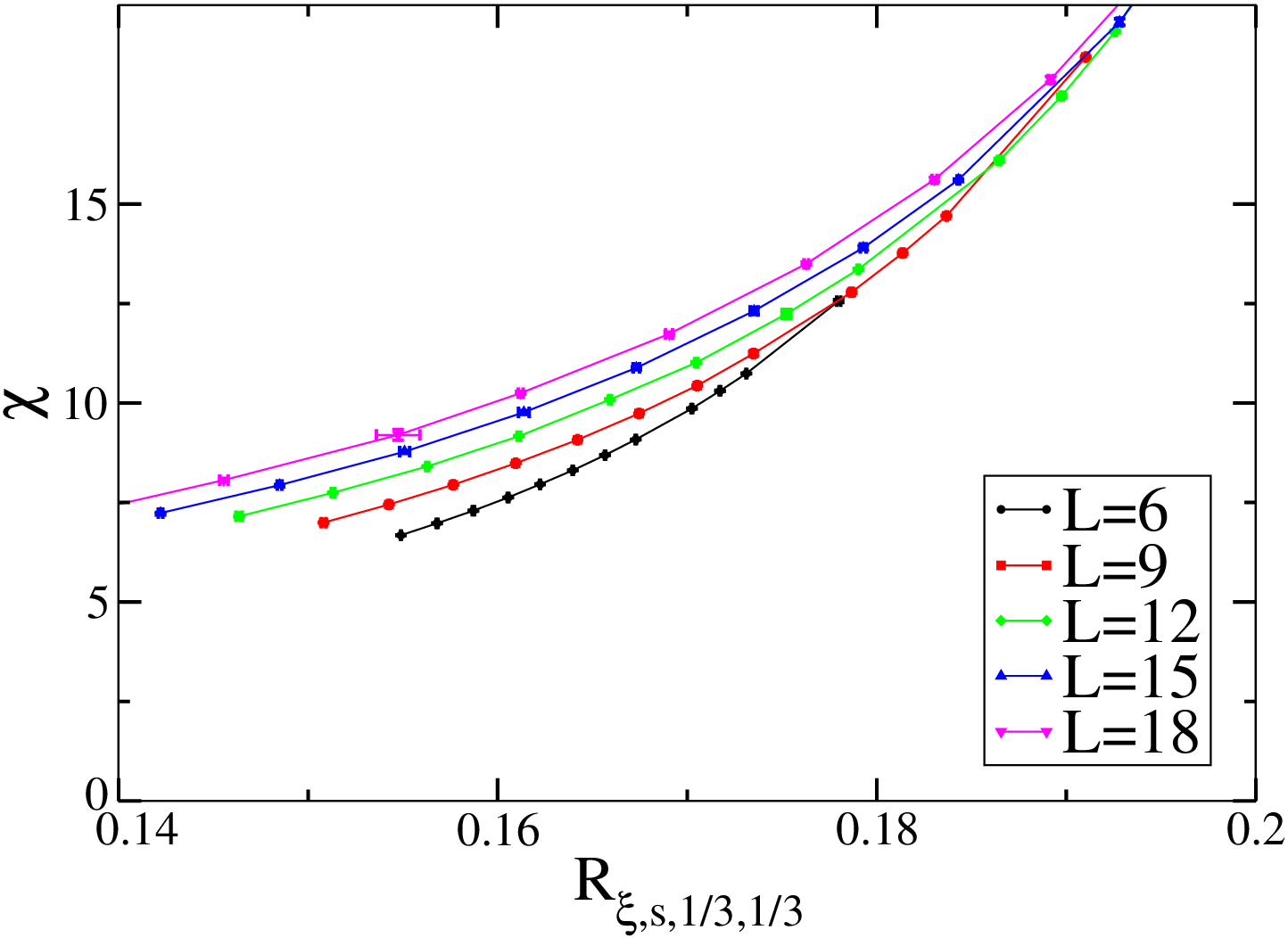}
\caption{(Color online) The two-point function at zero momentum $\chi$ as a
  function of the RG-invariant observable $\Rxist$. Lines are guides to the eye.}
\label{chi_vs_sumxil_hubbard}
\end{figure}

The exponent $\eta$ can be determined by analyzing the FSS behavior of
$\chicl$. To avoid using the values of $U_c$ and $\nu$ determined above, we
invert Eq.~(\ref{RG-scaling}) to obtain the scaling
variable $w$ as a function of $R$. Then, Eq.~(\ref{chifss}) can be rewritten
as
\begin{equation}
\chicl(R,L)=L^{2-z-\eta}f_{\chicl,R}(R),
\label{chi_vs_R}
\end{equation}
where corrections to scaling have been neglected. Since the previous analysis
has shown that $\Rxist$ is affected by small scaling corrections, we 
chose to analyze $\chicl$ using $R=\Rxist$. In
Fig.~\ref{chi_vs_sumxil_hubbard} we show $\chicl$ as a function of
$\Rxist$. The fact that $\chicl$
slowly grows with $L$ suggests a small value of the exponent $2-z-\eta$
that appears in Eq.~(\ref{chi_vs_R}).

For a quantitative analysis of the exponent $\eta$ we fitted
$\chicl(R,L)$ to a Taylor expansion of the right-hand side of Eq.~(\ref{chi_vs_R}),
using the QMC data for which $\Rxist\in [0.151,0.171]$;
for the central lattice size $L=12$, this interval in $\Rxist$ corresponds to
the range $U\in [3.6,4]$ that we used in the analysis of the $\nu$
exponent.  We performed a fit of the data for $\chicl(U,R)$ to
\begin{equation}
\chicl(R,L)=L^{1-\eta'}\sum_{n=0}^{n_{\rm max}}a_nR^n,\quad \eta'\equiv\eta+z-1,
\label{chi_vs_R_Taylor}
\end{equation}
with $R=\Rxist$ and leaving $\eta'$ and $\{a_n\}$ as free parameters. In
Eq.~(\ref{chi_vs_R_Taylor}) we have introduced for convenience the exponent
$\eta'$, which is defined such that $\eta'=\eta$ if $z=1$. In Table~\ref{fit_chi_vs_R_hubbard},
we report the fit results as a function of $n_{\rm max}$ and the minimum lattice size $L_{\rm min}$ taken into account. We
observe that $\chi^2/\DOF$ substantially decreases upon increasing the
expansion order from $n_{\rm max}=1$ to $n_{\rm max}=2$, while no appreciable
difference is found upon further increasing $n_{\rm max}$ to $n_{\rm max}=3$. Clearly, a parabolic approximation $n_{\rm max}=2$ is sufficient to
describe our MC data in the interval $\Rxist\in [0.151,0.171]$. On the other hand,
the $\chi^2/\DOF$ is large and acquires a small value for $L_{\rm min}=12$
only. This shows that scaling corrections give an important
contribution. Indeed, the fits indicate a value $\eta'\approx 0.7$: for such a value of $\eta'$ the
background contribution to $\chicl$ results in corrections to scaling with a
rather small exponent $\omega=1-\eta'\approx 0.3$.

\begin{table}
\caption{Results of the fit of $\chicl$ for the honeycomb Hubbard model to
  Eq.~(\ref{chi_vs_R_Taylor}) (first three sets) and to
  Eq.~(\ref{chi_vs_R_Taylor_background}) (last set), with $R=\Rxist$ and
  $\Rxist\in [0.151,0.171]$. The critical exponent $\eta'$ is defined as
  $\eta'\equiv\eta+z-1$, such that $\eta'=\eta$ if $z=1$. $L_{\rm min}$ is
  the minimum lattice size taken into account.}
\begin{ruledtabular}
\begin{tabular}{r@{\hspace{1em}}l@{}@{}.l@{}.@{}r}
 & \multicolumn{1}{c}{$L_{\rm min}$} & \multicolumn{1}{c}{$\eta'$} & \multicolumn{1}{c}{$\chi^2/\DOF$} \\
\hline
\multirow{3}{*}{$n_{\rm max}=1$} & $6$    & 0.771(2) & 1759.7/23 \\
& $9$    & 0.746(4) & 445.0/14 \\
& $12$   & 0.759(9) & 65.4/8 \\
\\
\multirow{3}{*}{$n_{\rm max}=2$} & $6$    & 0.766(2) & 1264.3/22 \\
& $9$    & 0.746(4) & 66.8/13 \\
& $12$   & 0.746(9) & 6.8/7 \\
\\
\multirow{3}{*}{$n_{\rm max}=3$} & $6$    & 0.765(2) & 1241.7/21 \\
& $9$    & 0.746(4) & 66.6/12 \\
& $12$   & 0.746(9) & 6.8/6 \\
\\
$n_{\rm max}=2$ & $6$    & 0.57(4)  &  112.7/20 \\
$m_{\rm max}=1$ & $9$    & 0.70(15) &  6.2/11 \\
\end{tabular}
\end{ruledtabular}
\label{fit_chi_vs_R_hubbard}
\end{table}

We thus consider the presence of an analytical background and fit our data to
\begin{equation}
\chicl(R,L)=L^{1-\eta'}\sum_{n=0}^{n_{\rm max}}a_nR^n+\sum_{m=0}^{m_{\rm max}}b_mR^m,\quad \eta'\equiv\eta+z-1,
\label{chi_vs_R_Taylor_background}
\end{equation}
with $R=\Rxist$. In Table~\ref{fit_chi_vs_R_hubbard}, we also report the fit
results with $n_{\rm max}=2$ and $m_{\rm max}=1$ for different 
$L_{\rm min}$. While the fit done using all
the available lattices shows a large $\chi^2/\DOF$, indicating the presence
of additional scaling corrections beyond those taken into account in
Eq.~(\ref{chi_vs_R_Taylor_background}), a good $\chi^2/\DOF$ is found for
$L_{\rm min}=9$. The fitted value of $\eta'$ is in full agreement with the results of the
fits to Eq.~(\ref{chi_vs_R_Taylor}) given in Table~\ref{fit_chi_vs_R_hubbard} (above); its error bar, which
is significantly larger than the one of the values obtained by the fits to Eq.~(\ref{chi_vs_R_Taylor})
gives a measure of the influence of the
slowly-decaying scaling corrections due to the background contribution that
is neglected in the fits to Eq.~(\ref{chi_vs_R_Taylor}). Moreover, the fitted
value of $\eta'$ for $L_{\rm min}=9$ agrees with the corresponding result for
$L_{\rm min}=6$.  Accordingly, we can regard the fit results for $L_{\rm
  min}=9$ with its uncertainty as a safe determination of the $\eta'$
exponent. We thus quote as a final result:
\begin{equation}
\begin{split}
\eta'=\eta+z-1=0.70(15),\\
\eta=0.70(15) \quad \text{(if $z=1$).}
\end{split}
\label{eta_hubbard}
\end{equation}

The estimate of Eq.~(\ref{eta_hubbard}) implies that the analytical part of
the free energy gives rise to slowly-decaying scaling corrections with an
effective correction-to-scaling exponent $\omega=2-z-\eta=0.30(15)$. In view
of the relatively small available lattice sizes, we repeated the FSS
analysis of $\Rxist$, this time including scaling corrections, with the aim
of checking the reliability of the estimates for $U_c$, $\nu$, and $\Rxist^*$
obtained above by neglecting scaling corrections. Indeed, even if
the RG-invariant observable $\Rxist$ appears to show small scaling
corrections, such a small value of $\omega$ may give rise to a drift in the
estimates of the critical parameters that is larger than the statistical error bar. We fitted $\Rxist$ to a Taylor expansion of Eq.~(\ref{RG-scaling}):
\begin{equation}
\begin{split}
R = R^* &+ \sum_{n=1}^{n_{\rm max}}a_n(U-U_c)^nL^{n/\nu} \\
        &+L^{-\omega}\sum_{m=0}^{m_{\rm max}}b_m(U-U_c)^mL^{m/\nu}.
\end{split}
\label{RG-scaling-exp-omega}
\end{equation}
In Table~\ref{fit_sumxil_hubbard}, we also report the fit results obtained for
fixed $\omega=0.15$, $0.3$, $0.45$, which reveal that the fitted value of $\nu$ is stable and in perfect agreement
with the estimate obtained by neglecting scaling corrections. However, we
observe that $U_c$ and $\Rxist^*$ exhibit a deviation with respect to the
previously obtained values $U_c=3.793(5)$, $\Rxist^*=0.1608(2)$. The
variation in $U_c$ is rather small, but larger than the error bars, whereas
the critical-point value $\Rxist^*$ exhibits a larger variation. Indeed,
residual scaling corrections affect in a statistically significant way the
fitted values of $U_c$ and $\Rxist^*$. Therefore, we choose more
conservative error bars for $U_c$ and $\Rxist^*$, which take into account
the results of Table~\ref{fit_sumxil_hubbard}, with and without considering
corrections to scaling. We obtain the estimates
\begin{equation}
\label{Uc_hubbard}
U_c=3.80(1),
\end{equation}
\begin{equation}
\label{nu_hubbard}
\nu=0.84(4),
\end{equation}
\begin{equation}
\label{sumxilstar_hubbard}
\Rxist^*=0.166(5).
\end{equation}
The final estimate for $U_c$ is also in full agreement with the less precise
estimate obtained by extrapolating the pseudocritical coupling
$U_{c,\Rxi}(L)$ (see Fig.~\ref{cross_hubbard}).

As a further check of the results presented in this section, we performed an additional FSS analysis of $\chicl$ as a function of $U$ and $L$, as done for the RG-invariant quantity $\Rxist$. The corresponding results are presented in Appendix \ref{sec:fsschicl} and corroborate the reliability of the obtained estimates.

\subsection{Kane-Mele-Hubbard model}\label{sec:results:km}

\begin{figure}
\includegraphics[width=\linewidth,keepaspectratio]{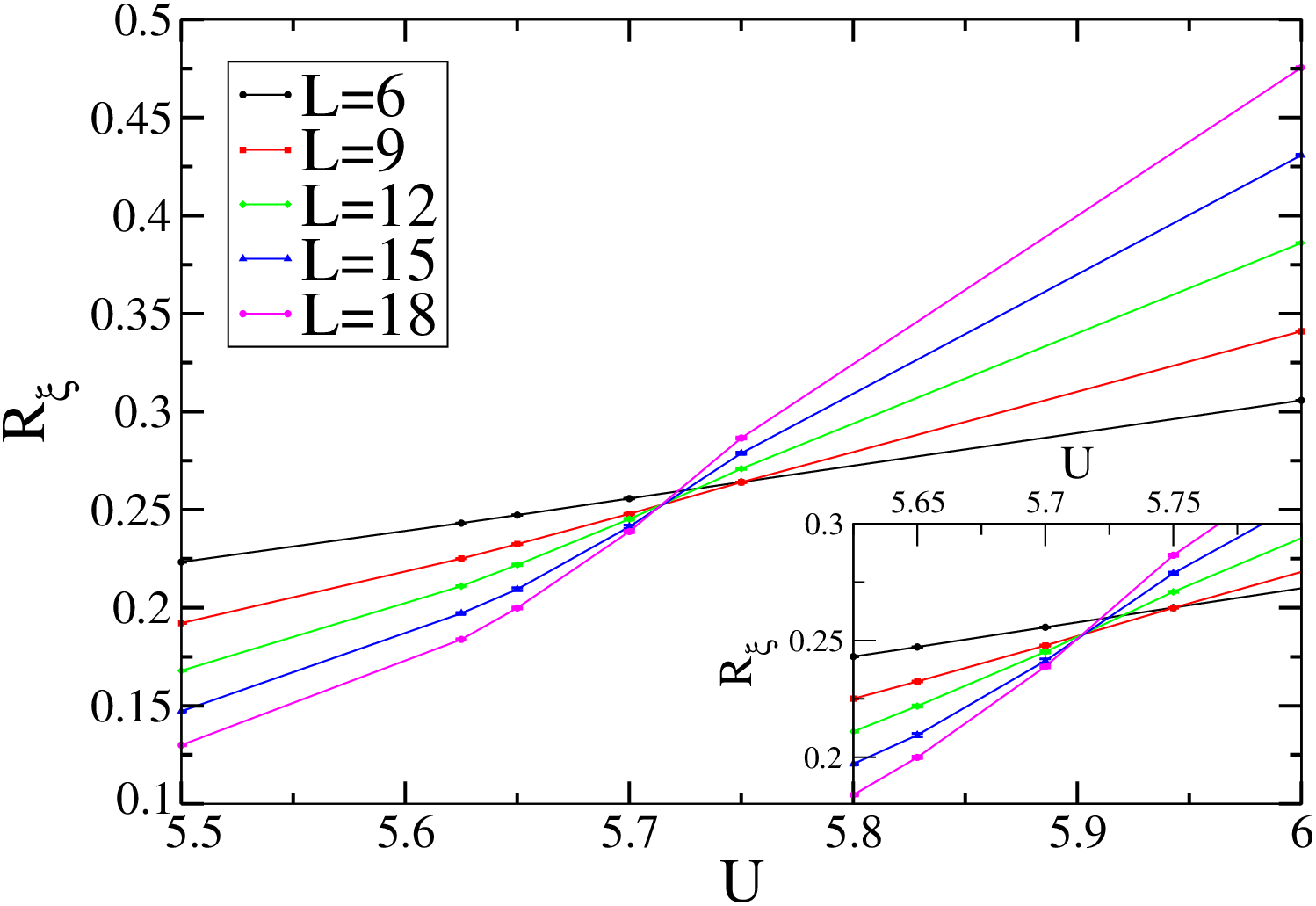}
\caption{(Color online) RG-invariant quantity $\Rxi$ for the
  Kane-Mele-Hubbard model. Lines are
  guides to the eye. Inset: magnification of the data
  close to their crossing at $U\approx 5.7$.}
\label{xil_km}
\end{figure}

\begin{figure}
\vspace{2em}
\includegraphics[width=\linewidth,keepaspectratio]{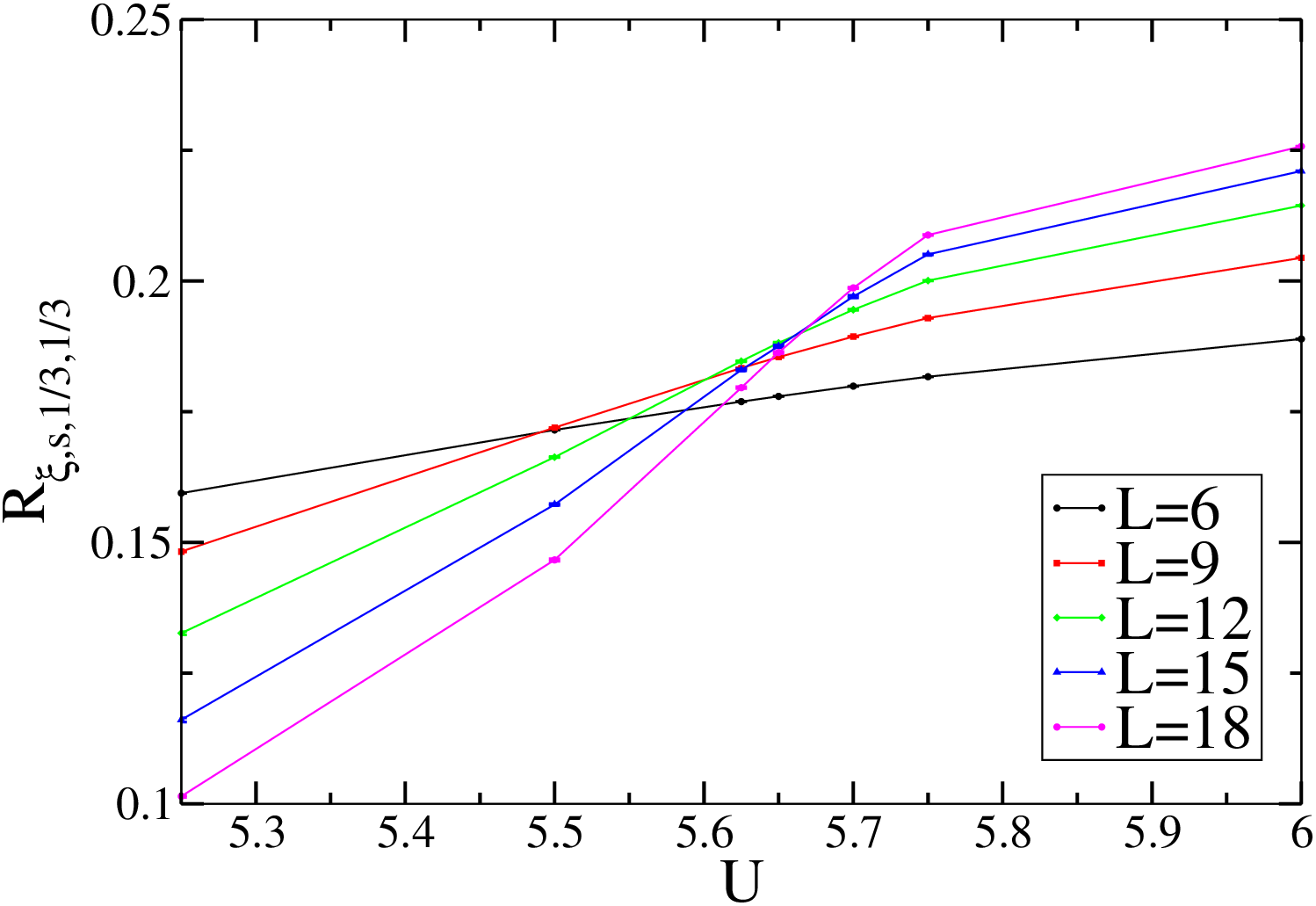}
\caption{(Color online)  Same as Fig.~\ref{xil_km} for $\Rxist$.}
\label{sumxil_km}
\end{figure}

\begin{figure}
\includegraphics[width=\linewidth,keepaspectratio]{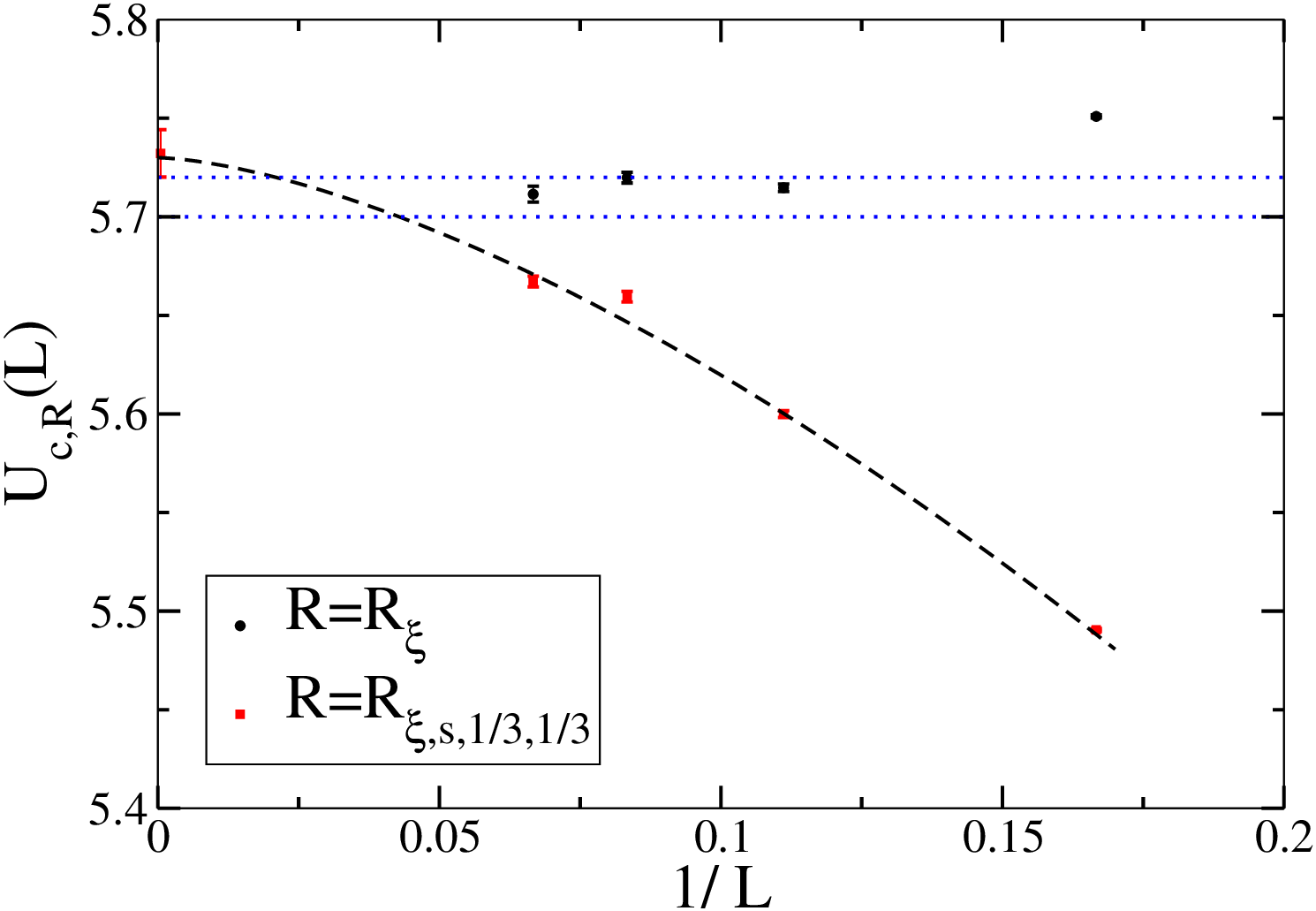}
\caption{(Color online)  Same as Fig.~\ref{cross_hubbard} for the
  Kane-Mele-Hubbard model. The plotted value of $U_{c,\Rxist}=5.73(1)$ for
  $L\rightarrow\infty$ has been obtained by fitting the data to
  Eq.~(\ref{pseudoconv}). The dashed line represents the right-hand side of
  Eq.~(\ref{pseudoconv}), with central values of the fit $U_c=5.73$. The
  dotted lines indicate the interval in the final estimate of the critical
  coupling $U=5.71(1)$ as reported in Eq.~(\ref{Uc_km}).}
\label{cross_km}
\end{figure}

We simulated the Kane-Mele-Hubbard model for lattice sizes $L=6$, $9$,
$12$, $15$, and $18$, setting the spin-orbit coupling $\lambda=0.2$. In Fig.~\ref{xil_km} (Fig.~\ref{sumxil_km}) we show
the RG-invariant quantity $\Rxi(U,L)$ [$\Rxist(U,L)$]
as a function of $U$ and for different lattice sizes $L$.
We observe that the curves of
$\Rxi(U,L)$ for $L\ge 9$ show a common intersection point at $U\approx 5.71$,
whereas the data for $\Rxist(U,L)$ exhibit a systematic drift of the
intersection point from $U\approx 5.5$ (the crossing point of the curves for
$L=6$ and $L=9$) towards larger values of $U$. In Fig.~\ref{cross_km}, we show
the pseudocritical couplings $U_{c,R}(L)$ as a function of the inverse
lattice size $L$, computed with the method mentioned in
Sec.~\ref{sec:results:hubbard}. Consistent with Figs.~\ref{xil_km}
and~\ref{sumxil_km}, $U_{c,\Rxi}(L)$ is constant within error bars for
$L\ge9$, while $U_{c,\Rxist}(L)$ increases with $L$. A fit of the results for
$U_{c,\Rxist}(L)$ to Eq.~(\ref{pseudoconv}) gives $U_c=5.73(1)$, with a large
$\chi^2/\DOF=22.5$. This suggests the presence of competing scaling
corrections in $\Rxist(U,L)$, which are not captured by
Eq.~(\ref{pseudoconv}). For this reason, the precision on the resulting value
of $U_c=5.73(1)$ has to be taken with caution, as it can be affected by a
systematic error. The limited lattice sizes available do not allow us to
further investigate the reliability of this result. Our final estimate of
$U_c$ is based on the FSS analysis of $\Rxi$ (see following). In
Fig.~\ref{cross_km}, we also show the right-hand side of Eq.~(\ref{pseudoconv}), as
fitted using the data for $U_{c,R}(L)$ with $R=\Rxist$. In line with the
considerations on the presence of a superposition of corrections to scaling,
some data points show a significant deviation from the
fitted curve.

\begin{table*}
\caption{Results of the fits of $R=\Rxi$ to Eq.~(\ref{RG-scaling-exp}) (first
  three sets) and to Eq.~(\ref{RG-scaling-exp-omega}) (last two sets) for
  the Kane-Mele-Hubbard model, with $U\in[5.625,5.75]$. $L_{\rm min}$ is the
  minimum lattice size taken into account in the fits.}
\begin{ruledtabular}
\begin{tabular}{r@{\hspace{1em}}l@{}@{}.l@{}.@{}r}
& \multicolumn{1}{c}{$L_{\rm min}$} & \multicolumn{1}{c}{$U_c$} & \multicolumn{1}{c}{$\nu$} & \multicolumn{1}{c}{$\Rxi^*$} & \multicolumn{1}{c}{$\chi^2/\DOF$} \\
\hline
\multirow{4}{*}{$n_{\rm max}=1$} & $6$  & 5.7524(4) & 0.727(3)  & 0.26101(8) & 4280.6/16 \\
& $9$  & 5.7104(5) & 0.716(5)  & 0.2516(2)  & 264.4/12 \\
& $12$ & 5.711(1)  & 0.77(1)   & 0.2517(6)  & 217.5/8 \\
& $15$ & 5.713(3)  & 0.84(4)   & 0.254(2)   & 195.3/4 \\
\\
\multirow{4}{*}{$n_{\rm max}=2$} & $6$  & 5.7335(3) & 0.587(3)  & 0.26097(6) & 2555.9/15 \\
& $9$  & 5.7155(5) & 0.657(5)  & 0.2526(2)  & 16.5/11 \\
& $12$ & 5.7157(9) & 0.68(1)   & 0.2528(5)  & 6.1/7 \\
& $15$ & 5.716(2)  & 0.714(29) & 0.253(2)   & 2.9/3 \\
\\
\multirow{4}{*}{$n_{\rm max}=3$} & $6$  & 5.7315(4) & 0.615(3)  & 0.26072(7) & 2471.3/14 \\
& $9$  & 5.7147(6) & 0.647(6)  & 0.2525(2)  & 9.8/10 \\
& $12$ & 5.7155(9) & 0.672(15) & 0.2528(5)  & 5.4/6 \\
& $15$ & 5.716(2)  & 0.715(35) & 0.253(2)   & 2.9/2 \\
\\
$n_{\rm max}=2$ & $6$  &  5.6982(9) & 0.665(3) & 0.2258(7) & 140.0/14 \\
$m_{\rm max}=0$ & $9$  &  5.715(2)  & 0.659(6) & 0.251(3)  & 16.4/10 \\
$\omega=0.785$ & $12$ &  5.711(8)  & 0.69(2)  & 0.244(16) & 5.8/6 \\
\\
$n_{\rm max}=3$ & $6$  &  5.6976(9) & 0.649(4) & 0.2261(7) & 107.8/13 \\
$m_{\rm max}=0$ & $9$  &  5.715(2)  & 0.646(7) & 0.253(3)  & 9.7/9 \\
$\omega=0.785$ & $12$ &  5.712(8)  & 0.68(2)  & 0.246(17) & 5.2/5
\end{tabular}
\end{ruledtabular}
\label{fit_xil_km}
\end{table*}
In order to determine the critical exponent $\nu$ and the critical coupling
$U_c$, we analyzed the FSS behavior of $\Rxi$ which, in this case,
appears to have reduced scaling corrections. We restrict the analysis to the
interval $U\in[5.625,5.75]$, around the expected critical point $U_c\approx
5.7$, as inferred from the analysis of the pseudocritical couplings. In
Table~\ref{fit_xil_km}, we report the results of the fits of $\Rxi$ to
Eq.~(\ref{RG-scaling-exp}). We observe that $\chi^2/\DOF$ decreases
significantly when we increase $n_{\rm max}$ from $n_{\rm max}=1$ to $n_{\rm max}=2$, and only marginally when $n_{\rm max}$
is set to $n_{\rm max}=3$. Thus, a quadratic approximation should be adequate to describe the
data for $\Rxi$ in the interval $U\in[5.625,5.75]$.
The fits with $n_{\rm max}=2$ show large values of $\chi^2/\DOF$ for $L_{\rm min}=6$, indicating important scaling corrections, and still a somewhat
large value of $\chi^2/\DOF$ for $L_{\rm min}=9$, suggesting the presence of
residual scaling corrections for $L=9$. The $\chi^2/\DOF$ ratio is good for
$L_{\rm min}\ge 12$. The fitted values of $U_c$, $\nu$, and $R_\xi^*$ are
essentially stable for $L_{\rm min}\ge 9$. Upon conservatively judging the
variation of the fit results for $\nu$ as obtained by these fits, one can
extract an estimate $\nu=0.68(3)$. This value agrees with that of the 3D XY
UC, $\nu=0.6717(1)$ \cite{PhysRevB.74.144506} (see the discussion in
Sec.~\ref{sec:model}). In view of value of $\chi^2/\DOF$ for $L_{\rm min}=9$,
we repeated the analysis by including scaling corrections. Our data do
not allow an independent determination of the $\omega$
exponent. Nevertheless, since we expect that the critical behavior belongs to
the 3D XY UC and since our fits to Eq.~(\ref{RG-scaling-exp}) are consistent
with this picture, we fitted $\Rxi$ to Eq.~(\ref{RG-scaling-exp-omega}),
fixing $\omega$ to the value of the leading irrelevant operator for the 3D XY
UC, $\omega=0.785(20)$ \cite{PhysRevB.74.144506}. The corresponding fit
results are given in Table~\ref{fit_xil_km} where, for completeness, we
also report the results of fits to Eq.~(\ref{RG-scaling-exp-omega}) with
$n_{\rm max}=3$. The results of the fits do not change significantly upon
varying $\omega=0.785(20)$ within one error bar. For this reason, we report
the fit results obtained by fixing $\omega$ to its central value $\omega=0.785$.

The inclusion of a correction-to-scaling term in the fits results in a large
reduction of the $\chi^2/\DOF$ ratio for the fits with $L_{\rm min}=6$, whose
corresponding results align to those obtained with $L_{\rm min}\ge
9$. However, the $\chi^2/\DOF$ ratio for $L_{\rm min}=6$ is still large,
indicating the presence of subleading scaling corrections. For $L_{\rm min}\ge 9$, the fits to Eq.~(\ref{RG-scaling-exp-omega}) exhibit
$\chi^2/\DOF$ ratios that are comparable to those obtained without scaling corrections. In
particular, for $n_{\rm max}=2$ and $L_{\rm min}=9$ the fits to
Eq.~(\ref{RG-scaling-exp-omega}) still show a somewhat large $\chi^2/\DOF$
ratio, suggesting either the presence of residual scaling corrections that
are not taken into account by the present analysis, or that the Taylor
expansion with $n_{\rm max}=2$ does not describe the data for
$U\in[5.625,5.75]$ and $L\le 9$ in a fully reliable way. Nevertheless, the
fitted values of $U_c$, $\nu$, and $R_\xi^*$ are essentially stable for
$L_{\rm min}\ge 9$, and upon including a correction-to-scaling term in the
FSS analysis. By conservatively judging the fit results, we obtain the
estimates
\begin{equation}
\label{Uc_km}
U_c=5.71(1),
\end{equation}
\begin{equation}
\label{nu_km}
\nu=0.68(3),
\end{equation}
\begin{equation}
\label{sumxilstar_km}
\Rxi^*=0.250(6).
\end{equation}
The estimates for $U_c$ and $\nu$ have been chosen so to agree with the
results of Table~\ref{fit_xil_km} for $n_{\rm max}\ge 2$ and $L_{\rm min}=9$,
$12$, including a variation of one error bar, with and without taking into
account scaling corrections. They are also in agreement with the fit results
for $L_{\rm min}=15$. The estimate for $\Rxi^*$ has been chosen such that it
agrees with the results of the fits that neglect scaling corrections for
$n_{\rm max}\ge 2$ and $L_{\rm min}=9$, $12$,  and with the results of the
fits that consider scaling corrections for $n_{\rm max}\ge 2$ and $L_{\rm
  min}=9$, including a variation of one error bar. The quoted value of
$\Rxi^*$ is also in agreement with the central value of the fits for $n_{\rm
  max}\ge 2$, $m_{\rm max}=0$, and $L_{\rm min}=12$, and with the fits done
without taking into account corrections to scaling, for $L_{\rm min}\ge
15$. The final estimate of $U_c$ is only in marginal agreement with the
estimate obtained by a extrapolating the pseudocritical coupling
$U_{c,\Rxist}(L)$. Such a difference does not contradict the precision of our
final result for $U_c$ because, as discussed above, the extrapolation of
$U_{c,\Rxist}(L)$ may be affected by a systematic error.

\begin{table}
\caption{Results of the fit of $\chi$ for the Kane-Mele-Hubbard model to
Eq.~(\ref{chi_vs_R_Taylor}) (first three sets) and to Eq.~(\ref{chi_vs_R_Taylor-omega}) (last set), with $R=R_\xi$ and
$R_\xi \in [0.197,0.287]$. $L_{\rm min}$ is the minimum lattice
size taken into account.}
\begin{ruledtabular}
\begin{tabular}{r@{\hspace{1em}}l@{}@{}.l@{}.@{}r}
 & \multicolumn{1}{c}{$L_{\rm min}$} & \multicolumn{1}{c}{$\eta'$}          & \multicolumn{1}{c}{$\chi^2/\DOF$} \\
\hline
\multirow{3}{*}{$n_{\rm max}=1$} & $6$  & 0.003(1) & 305.5/17 \\
& $9$  & 0.059(4) & 10.1/12 \\
& $12$ & 0.071(9) & 5.5/8 \\
\\
\multirow{3}{*}{$n_{\rm max}=2$} & $6$  & 0.003(1) & 305.2/16 \\
& $9$  & 0.068(5) & 2.9/11 \\
& $12$ & 0.08(1)  & 0.22/7 \\
\\
\multirow{3}{*}{$n_{\rm max}=3$} & $6$  & 0.003(1) & 299.8/15 \\
& $9$  & 0.068(5) & 2.9/10 \\
& $12$ & 0.08(1)  & 0.17/6 \\
\\
$n_{\rm max}=2$ & $6$  & 0.087(8) & 194.9/15 \\
$m_{\rm max}=0$  & $9$  & 0.076(21)& 2.7/10\\
$\omega=0.785$ & & &
\end{tabular}
\end{ruledtabular}
\label{fit_chi_vs_R_km}
\end{table}
In Table~\ref{fit_chi_vs_R_km}, we report the results of the fits of $\chicl$
to Eq.~(\ref{chi_vs_R_Taylor}) for $R=\Rxi$. We restrict the analysis to the
interval $\Rxi \in [0.197,0.287]$, which for lattice sizes $L=9-15$
corresponds to the interval $U \in [5.625,5.75]$ that we used to analyze the
FSS behavior of $\Rxi$. We observe a small decrease of the $\chi^2/\DOF$
ratio when we increase the expansion order from $n_{\rm max}=1$ to $n_{\rm
  max}=2$, while no appreciable difference is found upon further increasing
$n_{\rm max}$ to $n_{\rm max}=3$. The fits for $L_{\rm min}\ge 9$ exhibit a
good $\chi^2/\DOF$ ratio, and the fitted value of $\eta'$ is stable upon
increasing $L_{\rm min}$ and $n_{\rm max}$. As done for the FSS analysis of
$\Rxi$, in order to monitor the role of the corrections to scaling, we
repeated the fits including a correction-to-scaling term. We fitted the
data of $\chicl$ to
\begin{equation}
\begin{split}
\chicl(R,L)=&L^{1-\eta'}\left(\sum_{n=0}^{n_{\rm max}}a_nR^n+L^{-\omega}\sum_{m=0}^{m_{\rm max}}b_mR^m\right),\\
&\eta'\equiv\eta+z-1,
\end{split}
\label{chi_vs_R_Taylor-omega}
\end{equation}
using $\omega=0.785$. By conservatively judging
the variation of the results in Table~\ref{fit_chi_vs_R_km}, we estimate
\begin{equation}
\begin{split}
\eta'&=\eta+z-1=0.075(20),\\
\eta&=0.075(20) \quad \text{(if $z=1$)},
\end{split}
\label{eta_km}
\end{equation}
where the error bar essentially includes the estimates of all the fits. This
value differs from the expected $\eta$ exponent of the
3D XY UC, $\eta=0.0381(2)$ \cite{PhysRevB.74.144506}. Although  
the difference is within two error bars, it suggests the presence of
residual scaling corrections that are not fully taken into account by the
present analysis.

\subsection{$\pi$-flux Hubbard model}\label{sec:results:piflux}

\begin{figure}
\includegraphics[width=\linewidth,keepaspectratio]{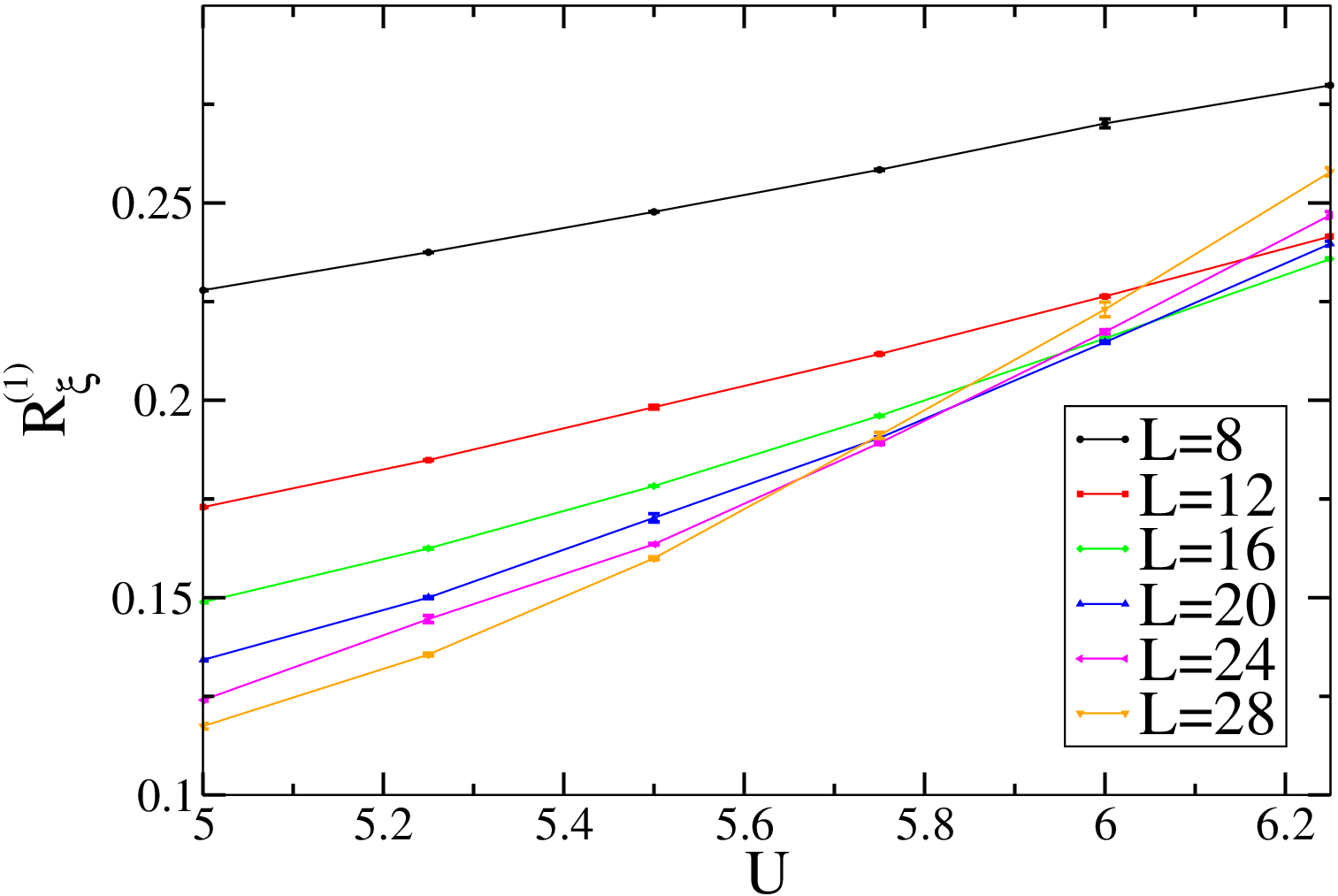}
\caption{(Color online) RG-invariant quantity $\Rxii{1}$ for the $\pi$-flux Hubbard
  model. Lines are guides to the eye.}
\label{xil1_piflux}
\end{figure}

\begin{figure}
\includegraphics[width=\linewidth,keepaspectratio]{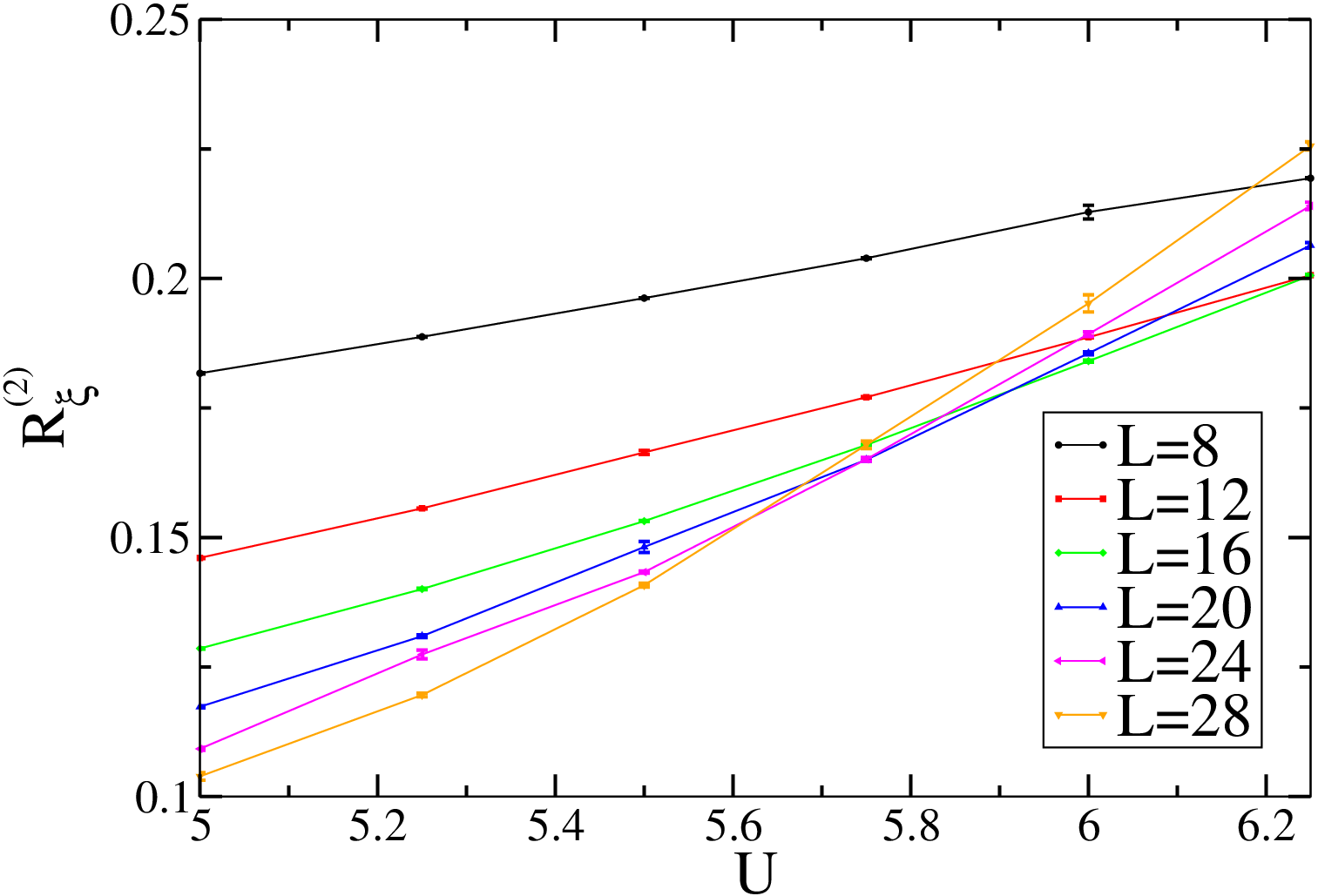}
\caption{(Color online) Same as Fig.~\ref{xil1_piflux} for $\Rxii{2}$.}
\label{xil2_piflux}
\end{figure}

\begin{figure}
\vspace{1em}
\includegraphics[width=\linewidth,keepaspectratio]{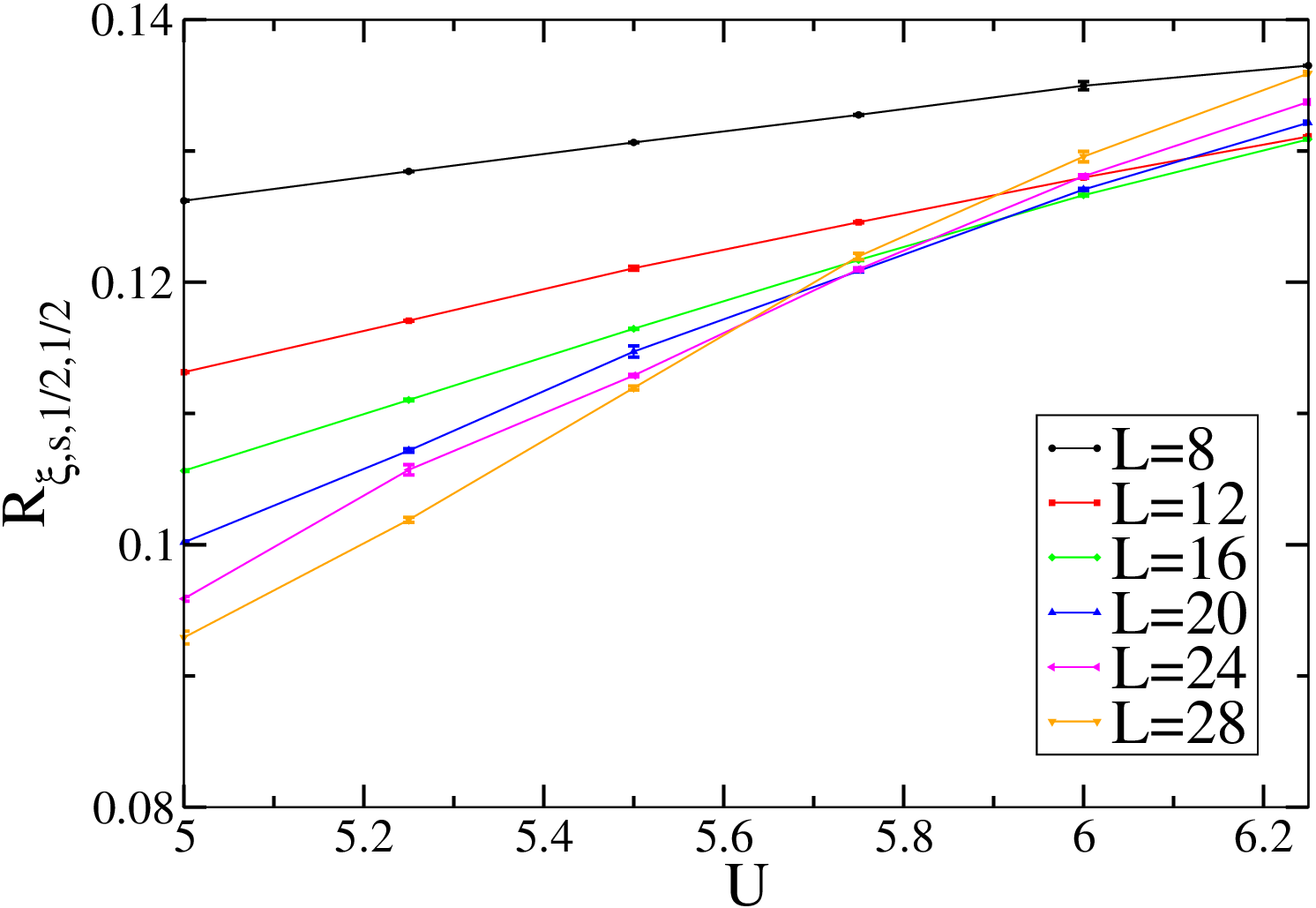}
\caption{(Color online) Same as Fig.~\ref{xil1_piflux} for $\Rxish$.}
\label{sumxil_shiftL2_piflux}
\end{figure}

\begin{figure}
\vspace{1em}
\includegraphics[width=\linewidth,keepaspectratio]{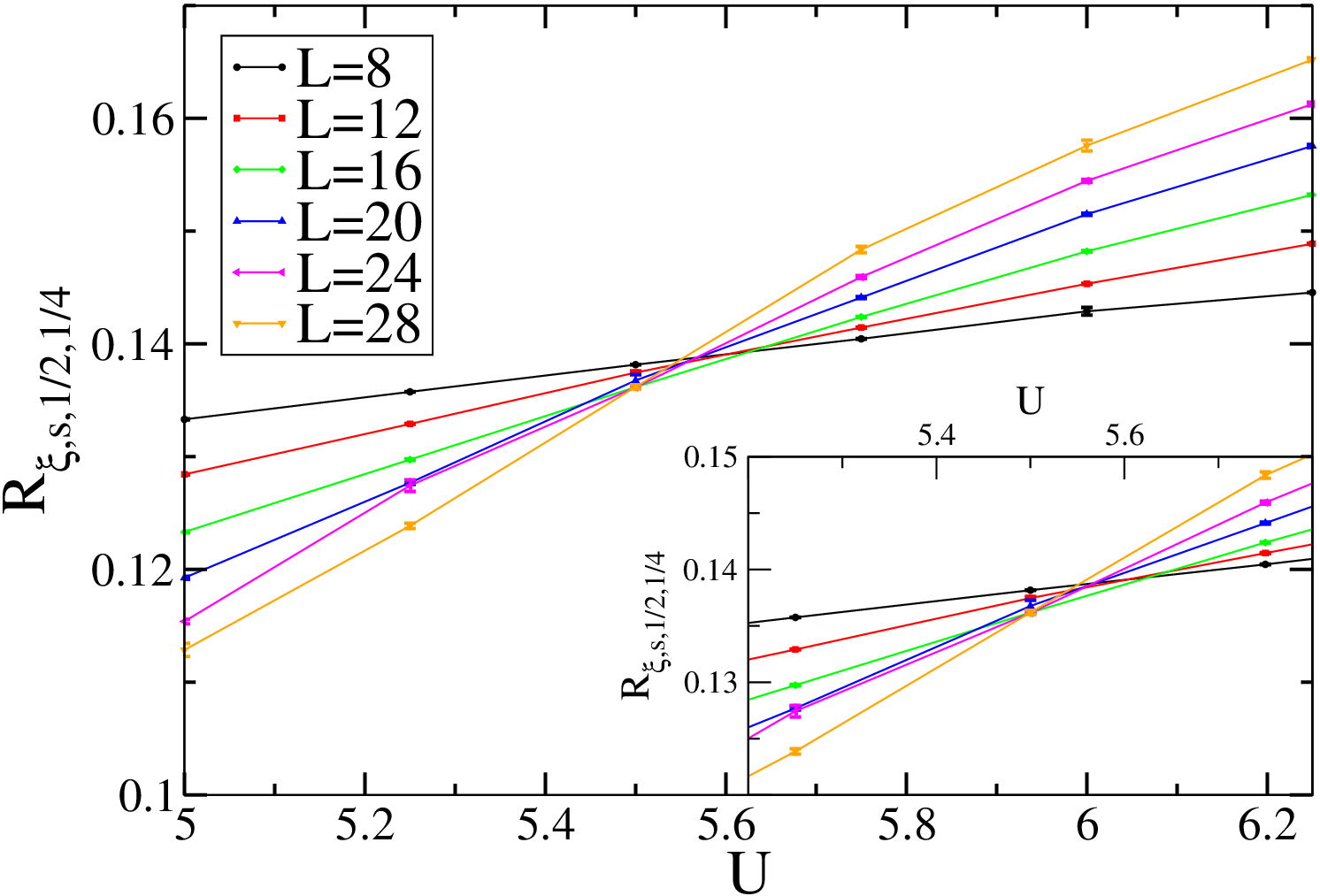}
\caption{(Color online) Same as Fig.~\ref{xil1_piflux} for $\Rxishq$. Inset: magnification of the data close to their crossing at $U\approx 5.5$.}
\label{sumxil_shiftL24_piflux}
\end{figure}

\begin{figure}
\includegraphics[width=\linewidth,keepaspectratio]{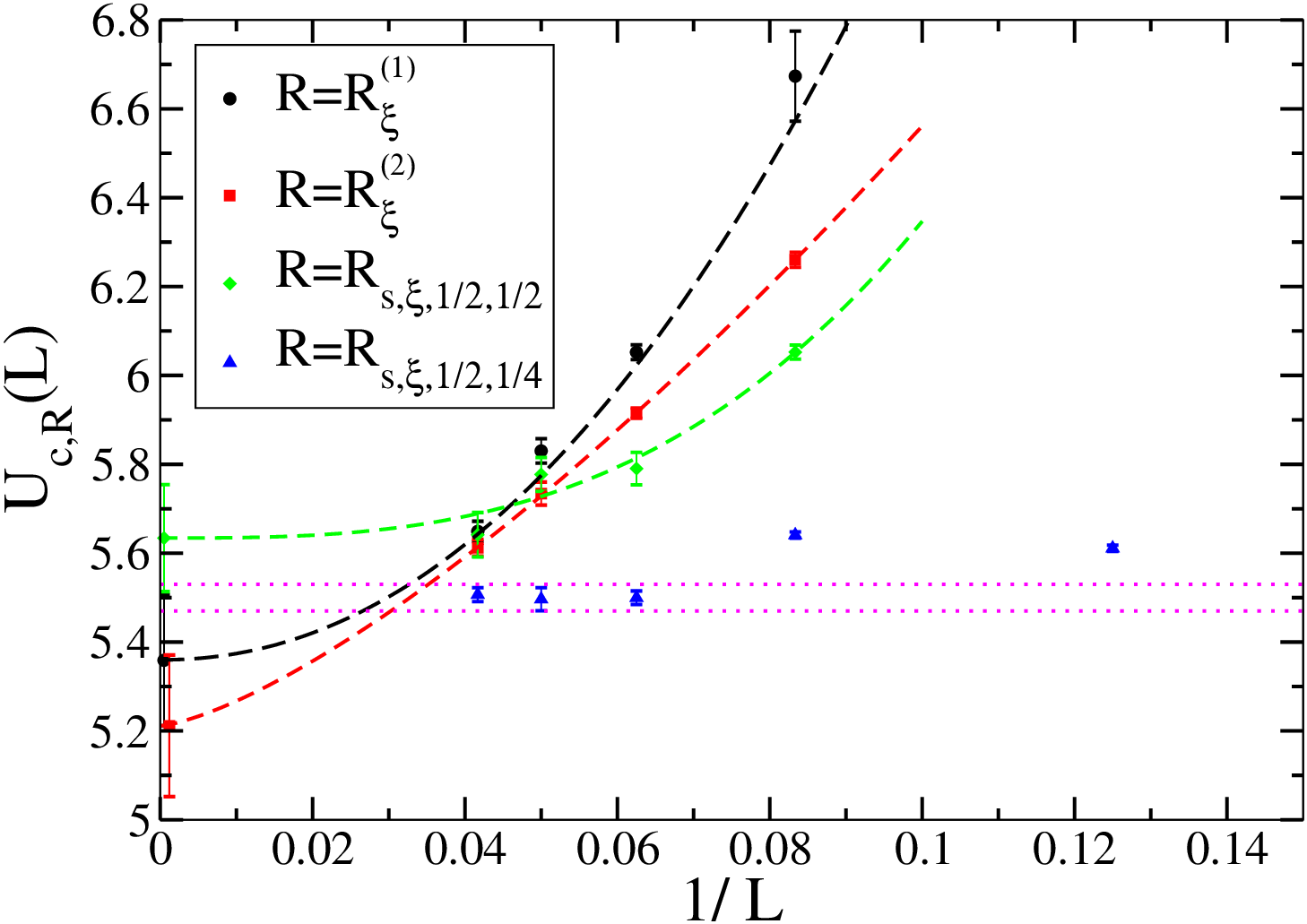}
\caption{(Color online) pseudocritical coupling $U_{c,R}$ for the $\pi$-flux Hubbard
  model and RG-invariant quantities $R=\Rxii{1}$, $\Rxii{2}$, $\Rxish$, and
  $\Rxishq$. The dashed lines represent the right-hand side of Eq.~(\ref{pseudoconv}),
  with the central values of the parameters as obtained by a fit to the
  right-hand side of Eq.~(\ref{pseudoconv}) and reported in
  Table~\ref{fit_pseudoconv_piflux}. For $R=\Rxii{1}$, $\Rxii{2}$, $\Rxish$,
  we also plot the extrapolated value of $U_{c,R}(L)$ for
  $L\rightarrow\infty$. The dotted lines indicate the interval in the final
  estimate of the critical coupling $U=5.50(3)$ as reported in
  Eq.~(\ref{Uc_piflux}).}
\label{cross_piflux}
\end{figure}

We carried out QMC simulations of the $\pi$-flux Hubbard model for
lattice sizes $L=8$, $12$, $16$, $20$, $24$, and $28$. In
Figs.~\ref{xil1_piflux}-\ref{sumxil_shiftL24_piflux} we show the RG-invariant quantities $\Rxii{1}$,
$\Rxii{2}$, $\Rxish$, and $\Rxishq$, respectively, as a function of $U$ and
for different lattice sizes $L$. Inspection of
Figs.~\ref{xil1_piflux}\,--\,\ref{sumxil_shiftL24_piflux} reveals that $\Rxii{1}$,
$\Rxii{2}$, $\Rxish$ are affected by significant scaling corrections, while
reduced corrections to scaling are observed for the RG-invariant observable
$\Rxishq$. This observation is confirmed by the analysis of the
pseudocritical couplings. In Fig.~\ref{cross_piflux}, we show 
$U_{c,R}(L)$ as a function of $1/L$, as obtained by numerically solving
Eq.~(\ref{crossdef}), with $R=\Rxii{1}$, $\Rxii{2}$, $\Rxish$, $\Rxishq$
and setting $c=4$. For the RG-invariant quantities $R=\Rxii{1}$, $\Rxii{2}$,
$\Rxish$, which exhibit significant scaling corrections, we fitted the
resulting pseudocritical couplings $U_{c,R}(L)$ for $L=12$, $16$, $20$, and
$24$ to Eq.~(\ref{pseudoconv}), leaving $U_c$, $A$, and $e$ as free
parameters. The fit results reported in Table~\ref{fit_pseudoconv_piflux} reveal a significant scatter in the
extrapolated $U_c$. Moreover, the $\chi^2/\DOF$ is in most cases large,
suggesting that these RG-invariant quantities are affected by a superposition
of competing scaling corrections that are not captured by
Eq.~(\ref{pseudoconv}) where only the leading scaling correction has been
taken into account. Moreover, for some of the RG-invariant observables
considered here, the crossing between the lattice sizes $L=12$ and $L=16$
lies outside the range of the available MC data. In this case, the
pseudocritical coupling has been obtained by extrapolating the values of
$R$; such a procedure may introduce a bias, which can contribute to the
observed spread in the extrapolated critical coupling $U_c$. The lack of
larger lattice sizes does not allow us to further investigate these
issues. On the other hand, the pseudocritical couplings $U_{c,R}(L)$ for
$R=\Rxishq$ appear to converge fast to $U_c$. Indeed, for $L\ge 16$,
$U_{c,R}(L)$ is stable within error bars, suggesting $U_c\simeq 5.5$.

\begin{table}
\caption{Results of fits of the pseudocritical couplings $U_{c,R}(L)$ to
  Eq.~(\ref{pseudoconv}) for the RG-invariant observables $R=\Rxii{1}$,
  $\Rxii{2}$, $\Rxish$.}
\begin{ruledtabular}
\begin{tabular}{l@{\hspace{2em}}.@{\hspace{2em}}.@{\hspace{2em}}.}
$R$        & \multicolumn{1}{c}{$U_c$} & \multicolumn{1}{c}{$e$} & \multicolumn{1}{c}{$\chi^2/\DOF$} \\
\hline
$\Rxii{1}$ & 5.36(15) & 2.1(6)   & 1.8 \\
$\Rxii{2}$ & 5.21(16) & 1.4(3)   & 0.05 \\
$\Rxish$   & 5.63(12) & 2.9(1.8) & 3.01 \\
\end{tabular}
\end{ruledtabular}
\label{fit_pseudoconv_piflux}
\end{table}

\begin{table*}
\caption{Same as Table~\ref{fit_sumxil_hubbard} for $R=\Rxishq$ and the
  $\pi$-flux Hubbard model, with $U\in[5.25,6]$.}
\begin{ruledtabular}
\begin{tabular}{r@{\hspace{1em}}l@{}@{}.l@{}.@{}r}
 & \multicolumn{1}{c}{$L_{\rm min}$} & \multicolumn{1}{c}{$U_c$} & \multicolumn{1}{c}{$\nu$} & \multicolumn{1}{c}{$\Rxishq^*$} & \multicolumn{1}{c}{$\chi^2/\DOF$} \\
\hline
\multirow{4}{*}{$n_{\rm max}=1$} & $8$  & 5.601(2) & 0.777(4) & 0.13899(3) & 976.5/20 \\
& $12$ & 5.561(3) & 0.836(7) & 0.13796(6) & 438.4/16 \\
& $16$ & 5.507(5) & 0.93(2)  & 0.1363(1)  & 117.2/12 \\
& $20$ & 5.50(1)  & 0.91(3)  & 0.1361(4)  & 88.9/11 \\
\\
\multirow{4}{*}{$n_{\rm max}=2$}  & $8$  & 5.592(2) & 0.768(4) & 0.13892(3) & 914.2/19 \\
& $12$ & 5.554(3) & 0.819(7) & 0.13792(6) & 383.0/15 \\
& $16$ & 5.495(5) & 0.888(14)& 0.1361(1)  & 22.7/11 \\
& $20$ & 5.49(1)  & 0.90(3)  & 0.1359(4)  & 21.4/7 \\
\\
\multirow{4}{*}{$n_{\rm max}=3$} & $8$  & 5.594(2) & 0.724(6) & 0.13890(3) & 842.0/18 \\
& $12$ & 5.556(3) & 0.782(9) & 0.13791(6) & 360.4/14 \\
& $16$ & 5.498(4) & 0.85(2)  & 0.1361(1)  & 16.7/10 \\
& $20$ & 5.49(1)  & 0.85(4)  & 0.1357(4)  & 16.1/6 \\
\end{tabular}
\end{ruledtabular}
\label{fit_sumxil_piflux}
\end{table*}

\begin{table*}
\caption{Same as Table~\ref{fit_sumxil_piflux} for $U\in [5.25,5.75]$ and $n_{\rm max}=1$.}
\begin{ruledtabular}
\begin{tabular}{l@{\hspace{4em}}.@{\hspace{4em}}.@{\hspace{4em}}.@{\hspace{4em}}.}
$L_{\rm min}$ & \multicolumn{1}{c}{$U_c$} & \multicolumn{1}{c}{$\nu$} & \multicolumn{1}{c}{$R_{\xi,s}^*$} & \multicolumn{1}{c}{$\chi^2/\DOF$} \\
\hline
$8$  &  5.596(2) &  0.765(6) &  0.13898(3) & 867.4/14 \\
$12$ &  5.556(3) &  0.806(9) &  0.13792(7) & 356.3/11 \\
$16$ &  5.503(4) &  0.87(2)  &  0.1362(1)  & 20.7/8 \\
$20$ &  5.49(1)  &  0.85(4)  &  0.1356(4)  & 14.7/5 \\
\end{tabular}
\end{ruledtabular}
\label{fit_sumxil_linear_piflux}
\end{table*}

Since the RG-invariant quantity $\Rxishq$ appears to have reduced scaling
corrections, we analyzed its FSS behavior to determine the
critical coupling $U_c$ and the exponent $\nu$. Similar to the
analysis in Secs.~\ref{sec:results:hubbard} and \ref{sec:results:km}, we
considered the QMC data in the interval $U\in[5.25,6]$ around the observed
common crossing of $\Rxishq$ at $U\simeq 5.5$ for $L\ge 16$. For this data set,
we fitted $\Rxishq$ to Eq.~(\ref{RG-scaling-exp}). In
Table~\ref{fit_sumxil_piflux}, we report the fit results for different
expansion orders $n_{\rm max}$ and minimum lattice sizes $L_{\rm min}$.

The ratio $\chi^2/\DOF$ decreases significantly
upon increasing $n_{\rm max}$ from $n_{\rm max}=1$ to $n_{\rm max}=2$, and
only marginally between $n_{\rm max}=2$ and  $n_{\rm max}=3$. This suggests that the Taylor expansion with $n_{\rm max}=2$
should be adequate in this interval of $U$. We find that
$\chi^2/\DOF$ decreases upon increasing $L_{\rm min}$, but remains
large even for the largest $L_{\rm min}$ used. This implies that, within the
available numerical precision, scaling corrections are important. The limited
number of data points does not allow for a more precise analysis, e.g., by
including corrections to scaling as done in Sec.~\ref{sec:results:hubbard}
(only four points are available for each $L$ in the chosen
interval). Nevertheless, Table~\ref{fit_sumxil_piflux} reveals that
for $n_{\rm max}\ge 2$, the fitted value of $U_c$ appears to be
stable for $L_{\rm min}\ge 16$, and the fitted exponent $\nu$ is essentially
in agreement with the estimate for the honeycomb Hubbard model, $\nu=0.84(4)$ [Eq.~(\ref{nu_hubbard})]. Similar results are found by analyzing the data in
a smaller interval $U\in [5.25,5.75]$ and setting $n_{\rm max}=1$. The
corresponding fit results are reported in Table~\ref{fit_sumxil_linear_piflux}.
Given the difficulty in studying the FSS of $\Rxishq$, we determined $U_c$ on the basis of the pseudocritical couplings $U_{c,R}(L)$ as
computed for $R=\Rxishq$. As mentioned above, $U_{c,R}(L)$ for $R=\Rxishq$ is
stable within error bars for $L\ge 16$: we find $U_{c,R}(L=16)=5.50(2)$,
$U_{c,R}(L=20)=5.50(3)$, $U_{c,R}(L=24)=5.51(2)$. Based on these values, we
arrive at the estimate
\begin{equation}
U_c=5.50(3)\,,
\label{Uc_piflux}
\end{equation}
where the error bar is chosen so that $U_c$ agrees with $U_{c,R}(L)$ for
$R=\Rxishq$ and $L\ge 16$, including a variation of one standard variation.

\begin{figure}
\vspace{2em}
\includegraphics[width=\linewidth,keepaspectratio]{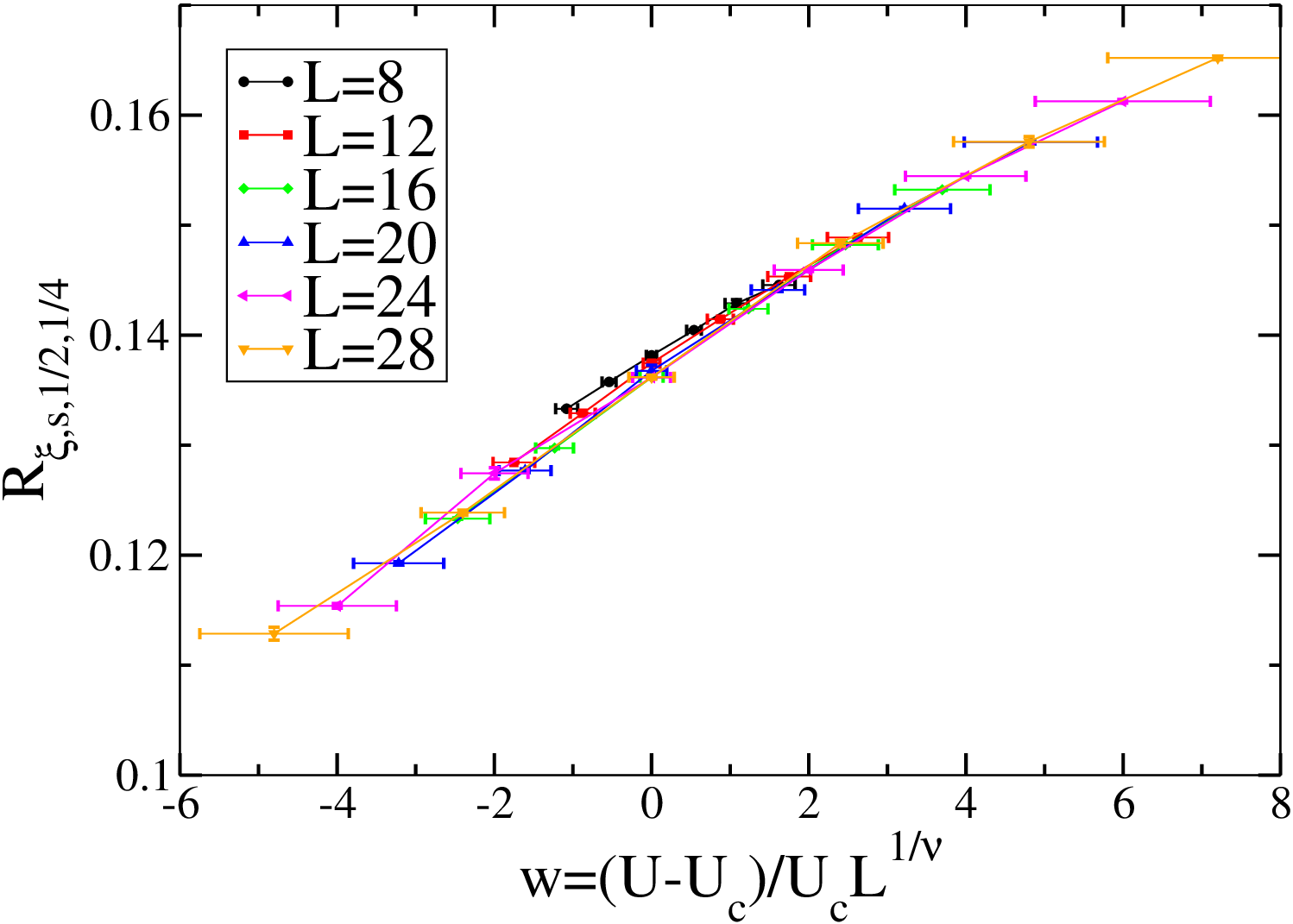}
\caption{(Color online) Scaling collapse for the RG-invariant quantity
  $\Rxishq$ for the $\pi$-flux Hubbard model. Lines are 
  guides to the eye. The scaling variable $w$ is
  computed using $U_c$ as given in Eq.~(\ref{Uc_piflux}) and $\nu$ as
  reported in Eq.~(\ref{nu_hubbard}).}
\label{collapse_piflux}
\end{figure}

To further strengthen the hypothesis that the critical behavior belongs to
the same UC as for the honeycomb Hubbard model, we produced a scaling
collapse for $\Rxishq$. Using the value of $U_c$ given in Eq.~(\ref{Uc_piflux}) and the
estimate of $\nu$ given in Eq.~(\ref{nu_hubbard}), we plot in
Fig.~\ref{collapse_piflux} $\Rxishq$ as a function of the scaling variable
$w$ defined in Eq.~(\ref{wdef}). Within the error bars, the data show a
collapse, consistent with the idea that the critical behavior belongs to the
Gross-Neveu-Heisenberg UC; the largest contribution to the
error bars on $w$ is due to the uncertainty on the exponent $\nu$, which is
responsible for the large error bars of the largest lattice sizes.

\begin{table}
\caption{Same as Table~\ref{fit_chi_vs_R_hubbard} for the $\pi$-flux Hubbard model for $R=\Rxishq$, with $\Rxishq\in[0.123,0.15]$.}
\begin{ruledtabular}
\begin{tabular}{r@{\hspace{1em}}l@{}@{}.l@{}.@{}r}
& \multicolumn{1}{c}{$L_{\rm min}$} & \multicolumn{1}{c}{$\eta'$}          & \multicolumn{1}{c}{$\chi^2/\DOF$} \\
\hline
\multirow{4}{*}{$n_{\rm max}=1$} & $8$  & 0.649(2) & 4373.9/23 \\
& $12$ & 0.681(3) & 2312.2/17 \\
& $16$ & 0.711(7) & 692.1/11 \\
& $20$ & 0.71(2)  & 80.3/6 \\
\\
\multirow{4}{*}{$n_{\rm max}=2$}  & $8$  & 0.679(2) & 768.4/22 \\
& $12$ & 0.670(3) & 239.5/16 \\
& $16$ & 0.696(7) & 43.0/10 \\
& $20$ & 0.70(2)  & 2.1/5 \\
\\
\multirow{4}{*}{$n_{\rm max}=3$} & $8$  & 0.679(2) & 765.9/21 \\
& $12$ & 0.668(4) & 236.6/15 \\
& $16$ & 0.697(7) & 30.3/9 \\
& $20$ & 0.71(2)  & 0.23/4 \\
\\
$n_{\rm max}=2$ & $8$  & 0.92(2)  & 104.9/20 \\
$m_{\rm max}=1$ & $12$ & 1.14(7)  & 32.6/14 \\
& $16$ & 0.99(20) & 7.7/8 \\
\end{tabular}
\end{ruledtabular}
\label{fit_chi_vs_R_piflux}
\end{table}

In Table~\ref{fit_chi_vs_R_piflux}, we report the results of fits of
$\chicl$ to Eq.~(\ref{chi_vs_R_Taylor}) for $R=\Rxishq$, in the interval
$\Rxishq\in[0.123,0.15]$ corresponding to
$U\in[5.25,5.75]$ for $L\ge 20$, to $U\in[5,6]$ for $L=16$, and to
$U\in[5,6.25]$ for $L\le 12$. We observe that $\chi^2/\DOF$ decreases
significantly between $n_{\rm max}=1$ and $n_{\rm max}=2$, while a
much smaller change is found between $n_{\rm max}=2$  and $n_{\rm max}=3$.
The value of $\chi^2/\DOF$ decreases upon disregarding
the smallest lattice size, but remains large even for $L_{\rm min}=20$,
signaling the importance of scaling corrections. Indeed, the fitted value of
$\eta'$ is large, $\eta'\sim 0.7$, which, analogous to the honeycomb Hubbard
model, implies the presence of slowly-decaying scaling corrections (compare with
Table~\ref{fit_chi_vs_R_hubbard}). As for the honeycomb Hubbard model, we
attempted to take into account these scaling corrections by including a
background term. The results of a fit of $\Rxishq$ to
Eq.~(\ref{chi_vs_R_Taylor_background}) using $n_{\rm max}=2$ and $m_{\rm max}=1$ are given in
Table~\ref{fit_chi_vs_R_piflux}. The fitted values of $\eta'$ do not exhibit
stability, and a small value of $\chi^2/\DOF$ is found for $L_{\rm min}=16$
only; in this case the fitted value of $\eta'$ agrees within error bars with
the estimate for the honeycomb Hubbard model [Eq.~(\ref{eta_hubbard})]. The
available data points do not allow for a more detailed
analysis. Nevertheless, there is little doubt that $\eta'$ (and hence $\eta$,
assuming $z=1$) is large, consistent with the Gross-Neveu-Heisenberg UC.

\section{Summary}\label{sec:summary}

We investigated the critical behavior of the honeycomb and the $\pi$-flux Hubbard model,
as well as the Kane-Mele-Hubbard model. Our main findings are as follows.

(i) By means of a FSS analysis that exploits RG-invariant observables, we
  determined the value of the critical coupling [Eq.~(\ref{Uc_hubbard}))] and
  an estimate of the critical exponents $\nu$ [Eq.~(\ref{nu_hubbard})] and
  $\eta$ [Eq.~(\ref{eta_hubbard})] for the Hubbard model on the honeycomb
  lattice (see Sec.~\ref{sec:results:hubbard}). The critical exponents are
  consistent with Gross-Neveu-Yukawa theory, in particular with a summation
  of the $\varepsilon$-expansion to the first loop that gives $\nu=97/110\simeq 0.88$, $\eta=0.8$.
  This justifies {\it a posteriori} the use of these critical
  exponents to obtain a scaling collapse in a previous QMC study of the
  honeycomb Hubbard model \cite{Assaad13},
  and of the Kane-Mele-Coulomb model \cite{PhysRevB.90.085146} for which
  the long-range Coulomb repulsion is expected to be marginally irrelevant \cite{Herbut01}.
  On the other hand, our determination of the critical exponents is not compatible
  with recent functional RG results \cite{PhysRevB.89.205403}.
  Our $U_c$ is in line with the value $U_c\simeq 3.78$
  reported in Ref.~\cite{Assaad13}.

(ii) Most notably, the critical behavior of the Hubbard model on the
  honeycomb lattice is characterized by a large value of the $\eta$
  exponent. As a consequence, the singular part of the two-point function of
  the order parameter decays fast as a function of the distance, so that the
  short-distance nonuniversal behavior gives a significantly large
  contribution to the spatial correlations. This results in slowly-decaying
  corrections to scaling that originate from the analytic part of the
  free energy and are characterized by a small effective
  correction-to-scaling exponent $\omega=0.30(15)$ [see the discussion after
  Eq.~(\ref{eta_hubbard})].  For comparison, for 3D classical
  $O(N)$ models $\eta\lesssim 0.04$, so that the leading scaling correction
  is due to the leading irrelevant operator, with $\omega\approx 0.8$
  \cite{PV-02}. Examples of classical models affected by slowly decaying
  scaling corrections are the 3D site-dilute and bond-dilute
  Ising models, where $\omega=0.33(3)$ \cite{HPPV-07}; for this UC the
  currently most precise critical exponents were
  obtained by simulating a classical 3D spin model with a
  lattice size up to $L=192$ \cite{HPPV-07}.  The presence of slowly-decaying
  scaling corrections in the Gross-Neveu-Heisenberg UC hinders
  a precise determination of the exponent $\eta$.

(iii) We analyzed the critical behavior of the Kane-Mele-Hubbard model
  with spin-orbit coupling $\lambda=0.2$
  (see Sec.~\ref{sec:results:km}), including a determination  of the critical
  coupling [Eq.~(\ref{Uc_km})] and the critical exponents
  $\nu$ [Eq.~(\ref{nu_km})] and $\eta$ [Eq.~(\ref{eta_km})]. The analysis
  confirms that the critical behavior belongs to the 3D XY UC,
  whose critical exponents are $\nu=0.6717(1)$, $\eta=0.0381(2)$
  \cite{PhysRevB.74.144506}. For this UC, the leading corrections to scaling
  are due to the leading irrelevant operator, whose negative RG-dimension is
  $\omega=0.785(20)$ \cite{PhysRevB.74.144506}. Assuming that the realization
  of the 3D XY UC by the Kane-Mele-Hubbard model does not generate
  additional irrelevant operators with a smaller negative RG-dimension, 
  $\omega=0.785(20)$ \cite{PhysRevB.74.144506} should characterize
  the leading scaling corrections [cf. the Hubbard model, where
  $\omega=0.30(15)$, see discussion after Eq.~(\ref{eta_hubbard})]. Our
  analysis of the $\eta$ exponent shows a small deviation, less than two
  error bars, from the precise determination for the 3D XY UC
  $\eta=0.0381(2)$ \cite{PhysRevB.74.144506}, suggesting the presence of
  residual scaling corrections that are not fully taken into account by the
  present analysis.

(iv) We analyzed the critical behavior of the $\pi$-flux Hubbard model
  (see Sec.~\ref{sec:results:piflux}). Although the available MC data do not
  allow for an independent determination of the critical exponents, we 
  provided evidence that the critical behavior is consistent with the
  Gross-Neveu-Heisenberg UC.

(v) Using the notion of a pseudocritical coupling (cf. discussion at the
  end of Sec.~\ref{sec:fss}) we determined the value of the critical
  coupling $U_c$ [Eq.~(\ref{Uc_piflux})] for the $\pi$-flux Hubbard model. A
  comparison with the corresponding value for the Hubbard model shows an
  interesting relation between the two critical couplings. By rescaling the
  values of $U_c$ [Eqs.~(\ref{Uc_hubbard}) and (\ref{Uc_piflux})] with the
  geometric average of the velocities at the Dirac cones
  [Eq.~(\ref{velocities})], we obtain
  \begin{align}\nonumber
    &\frac{U_c}{\sqrt{v_xv_y}} \simeq 4.4 \quad (\text{honeycomb Hubbard model}),\\
    &\frac{U_c}{\sqrt{v_xv_y}} \simeq 4.2 \quad (\text{$\pi$-flux Hubbard model}).
  \label{ratioUcv}
  \end{align}
  These results suggest that the velocities at the Dirac cones are the main
  contribution to the renormalization of $U_c$. 
  Note that the bandwidth $W$ is similar (but not equal) in the two models:
  $W=6$ for the honeycomb Hubbard model, and $W=4\sqrt{2}\simeq 5.6$ for the
  $\pi$-flux Hubbard model \cite{IAS-14}. The residual difference in
  the ratios in Eq.~(\ref{ratioUcv}) may originate from the ratio of the two bandwidths.

(vi)
In this work, we studied the critical behavior of the magnetic order parameter only.
Recent studies of the honeycomb Hubbard model \cite{Assaad13} and of the $\pi$-flux Hubbard model \cite{IAS-14} provided evidence that the opening of the single particle gap coincides with the onset of antiferromagnetic order. Together with these results, our analysis supports the validity of the Gross-Neveu-Yukawa theory, which predicts that the fermionic and bosonic degrees of freedom become critical at the same value of $U_c$, resulting in a direct transition between a semimetallic phase and an antiferromagnetic state.

(vii) Our FSS analysis exploited RG-invariant observables defined as
  ratios $\xi/L$ of the finite-size correlation length $\xi$ and the system size
  $L$. In a finite system, there is no unique definition of $\xi$, and we 
  defined several correlation lengths that
  are inequivalent in the FSS limit (see Appendix~\ref{sec:xi}). This freedom
  in the definition of $\xi$ leads us to several RG-invariant observables, some of them approximately improved, i.e., showing
significantly reduced
  scaling corrections. Improved observables and improved
  models are instrumental in high-precision studies of critical phenomena
  \cite{PV-02}.

\begin{acknowledgments}
We thank the LRZ-M\"unich and the J\"ulich Supercomputing Centre for CPU time,
and acknowledge financial support from the Deutsche Forschungsgemeinschaft Grants No.
AS120/9-1 and Ho 4489/3-1 (FOR 1807). I.F.H. is supported by the NSERC of Canada.
We acknowledge support from the Max Planck Institute for the Physics of
Complex Systems in Dresden, where this work was initiated in the framework of
the Advanced Study Group 2012/2013.
We thank T. C. Lang for useful communications.
F.P.T. is grateful to A. W. Sandvik and E. Vicari for useful discussions. 
\end{acknowledgments}

\appendix

\section{Finite-size correlation length}\label{sec:xi}

\subsection{Regular lattices}\label{sec:xi:regular}

On an infinite lattice with dimension $d$, the second-moment correlation length $\xi$ is defined as
\begin{equation}
\xi^2 \equiv \frac{1}{2d}\frac{\sum_{\vec{x}}|\vec{x}|^2\C(\vec{x})}{\sum_{\vec{x}}\C(\vec{x})},
\label{xi2def}
\end{equation}
where the sum is over the points $\vec{x}$ on the lattice, $\C(\vec{x})$ is the two-point function of the order parameter, and $|\vec{x}|$ is the Euclidean length of the vector $\vec{x}$.
Here we assume that the order parameter is a local quantity defined in terms of the observables on a single lattice site $\vec{x}$.
Equation~(\ref{xi2def}) can be written as
\begin{equation}
\xi^2 = -\frac{1}{2d\widetilde{\C}(\vec{p}=0)}\sum_i\frac{\partial^2\widetilde{\C}(\vec{p})}{\partial p_i\partial p_i}\Big|_{\vec{p}=0},
\label{xi2inF}
\end{equation}
where $\widetilde{\C}(\vec{p})$ is the Fourier transform of $\C(\vec{r})$,
\begin{equation}
\widetilde{\C}(\vec{p}) \equiv \sum_{\vec{r}} e^{i\vec{p}\vec{r}}\C(\vec{r}),
\label{fourierdef}
\end{equation}
and the derivatives of $\widetilde{\C}(\vec{p})$ in Eq.~(\ref{xi2inF}) are taken with respect to the Euclidean basis, or with respect to another orthonormal basis.
In the following, we specialize the discussion to the case $d=2$, i.e., of a two-dimensional lattice. An extension to higher-dimensional lattices is straightforward.

In a finite lattice with size $L$ there is not a unique definition of $\xi$,
but, in the presence of periodic boundary conditions one can substitute the derivative in Eq.~(\ref{xi2inF}) with a finite
incremental ratio calculated on the smallest momentum of the lattice $p_{\rm min}\sim 1/L$. To this end, we first analyze the properties of the  Taylor expansion of $\widetilde{\C}(\vec{p})$ for $\vec{p}\rightarrow 0$
\footnote{Even if the Fourier transform $\protect\widetilde{C}(\vec{p})$ is not analytic, we can still regard the expansion of Eq.~(\ref{TaylorG}) as describing the small-momentum behavior of a system with a large but finite size $L$, where the smallest momentum of the lattice $p_{\rm min}\sim 1/L$. In fact, all we need for the FSS analysis is to provide a definition of $\xi$ such that the ratio $\xi/L$ is RG-invariant and the finite-size correlation length $\xi(L)$ is analytic in an interval around the critical point.}:
\begin{multline}
\widetilde{\C}(\vec{p}) = \widetilde{\C}(0) + g_xp_x + g_yp_y+ g_{xx}p_x^2 + g_{xy}p_xp_y \\+g_{yy}p_y^2 + O(p^4),
\label{TaylorG}
\end{multline}
where $p_x$, $p_y$ are the components of $\vec{p}$ in the Euclidean basis (in
general not coinciding with the reciprocal lattice basis). The symmetries of
the lattice constrain the coefficients $g_i$, $g_{ij}$ in
Eq.~(\ref{TaylorG}). In fact, the invariance under a rotation by an angle
$\theta$, described by
\begin{equation}
\begin{pmatrix} p_x \\ p_y \end{pmatrix}
\rightarrow
\begin{pmatrix}
\cos \theta & -\sin \theta\\
\sin \theta & \cos \theta
\end{pmatrix}
\begin{pmatrix} p_x \\ p_y \end{pmatrix}
\label{rotationsym}
\end{equation}
with $\theta\ne 0$, $\pi$ requires the coefficients to satisfy
\begin{equation}
g_x=g_y=g_{xy}=0, \quad g_{xx}=g_{yy}\equiv A,
\end{equation}
so that Eq.~(\ref{TaylorG}) can be simplified to
\begin{equation}
\widetilde{\C}(\vec{p}) = \widetilde{\C}(0) + A\left(p_x^2+p_y^2\right) + O(p^4).
\label{TaylorGsimplepxpy}
\end{equation}
Equation~(\ref{TaylorGsimplepxpy}) holds, in particular, for the square lattice ($\theta=\pi/2$) and for the triangular lattice ($\theta=2\pi/3$). By inserting Eq.~(\ref{TaylorGsimplepxpy}) in Eq.~(\ref{xi2inF}), we find ($d=2$)
\begin{equation}
\xi^2=-\frac{A}{\widetilde{\C}(0)},
\end{equation}
so that the expansion of Eq.~(\ref{TaylorGsimplepxpy}) can be expressed as
\begin{equation}
\widetilde{\C}(\vec{p}) = \widetilde{\C}(0)\left[1 -\xi^2 \left(p_x^2+p_y^2\right)\right] + O(p^4).
\label{TaylorGsimple}
\end{equation}
Then, for any function $\Delta(\vec{p})$ that has a Taylor expansion of the form
\begin{equation}
\Delta(\vec{p}) = p_x^2 + p_y^2 + O(p^4),
\label{Deltaexp}
\end{equation}
we find that
\begin{equation}
\frac{1}{\Delta(\vec{p})}\left[\frac{\widetilde{\C}(0)}{\widetilde{\C}(\vec{p})}-1\right] = \xi^2 + O(p^2),\qquad \vec{p}\rightarrow 0.
\label{finitediff2}
\end{equation}
This result suggests to define, on a {\it finite} lattice with size $L$, the correlation length $\xi(L)^2$ as
\begin{equation}
\xi(L)^2 \equiv \frac{1}{\Delta(\vec{p}_\text{min})}\left[\frac{\widetilde{\C}(0)}{\widetilde{\C}(\vec{p}_\text{min})}-1\right],
\label{xi2finite}
\end{equation}
where $\vec{p}_\text{min}$ is the minimum momentum on a lattice of size
$L$. In a two-dimensional lattice there are two such minimum momenta where,
by virtue of the lattice symmetry, $\widetilde{\C}(\vec{p})$ takes the same
value. For simplicity,  in Eq.~(\ref{xi2finite}), we neglected a possible
dependence of $\xi(L)$ on additional parameters of the model,
such as the Hubbard coupling $U$ or the temperature. A comparison of
Eq.~(\ref{finitediff2}) with Eq.~(\ref{xi2inF}) shows that for
$L\rightarrow\infty$ the finite-size correlation length $\xi(L)$ coincides
with the second-moment correlation length $\xi$ up to corrections of order
$\sim p_\text{min}^2\sim 1/L^2$.

The choice of $\Delta(\vec{p})$ to be used in Eq.~(\ref{xi2finite}) is
usually dictated by the solution of a Gaussian model on the same lattice. For
such a model the Fourier transform of the two-point function can be
determined as
\begin{equation}
\widetilde{\C}(\vec{p})=\frac{\widetilde{\C}(0)}{1 + \xi^2\Delta(\vec{p})},
\label{GaussG}
\end{equation}
where the function $\Delta(\vec{p})$ depends on the lattice, its
normalization is fixed by Eq.~(\ref{Deltaexp}) and, in agreement with
Eq.~(\ref{TaylorGsimple}), the coefficient in front of $\Delta(\vec{p})$ is
equal to the second-moment correlation length. Inverting Eq.~(\ref{GaussG}),
we find that for a Gaussian model $\xi$ is exactly given by
\begin{equation}
\xi^2 = \frac{1}{\Delta(\vec{p})}\left[\frac{\widetilde{\C}(0)}{\widetilde{\C}(\vec{p})}-1\right].
\label{xi2Gauss}
\end{equation}
For an interacting model on a finite regular lattice, we can use the
definition of Eq.~(\ref{xi2finite}) for the finite-size correlation length
$\xi(L)$ and replace $\Delta(\vec{p})$ with the function obtained for the
Gaussian model on the same lattice. With this choice, the definition of
Eq.~(\ref{xi2finite}) gives exactly the second-moment correlation length in
the case of a Gaussian model. A different choice
of $\Delta(\vec{p})$, with the same normalization of Eq.~(\ref{Deltaexp}),
would give rise to different corrections $\propto 1/L^2$, which are in any
case negligible compared to the leading scaling correction.

For a square lattice, the function $\Delta(\vec{p})$ is
\begin{equation}
\Delta(\vec{p}) = 4\left[\sin \left(\frac{p_x}{2}\right)^2 + \sin \left(\frac{p_y}{2}\right)^2\right].
\end{equation}
The direct lattice basis $\{\vec{a}_1,\vec{a}_2\}$ and the reciprocal one $\{\vec{b}_1,\vec{b}_2\}$ of the square lattice are
\begin{equation}
\vec{a}_1=\begin{pmatrix}1 \\ 0\end{pmatrix},  \ \vec{a}_2=\begin{pmatrix}0 \\ 1\end{pmatrix}, \quad \vec{b}_1=\begin{pmatrix}0 \\ 1\end{pmatrix},  \ \vec{b}_2=\begin{pmatrix}1 \\ 0\end{pmatrix},\\
\label{squarebasis}
\end{equation}
where the lattice constant has been set to $1$ and the basis has been normalized such that
\begin{equation}
\vec{a}_i\cdot\vec{b}_j=\delta_{ij}.
\label{basisnorm}
\end{equation}
On a finite lattice with size $L$, the two minimum momenta are $\vec{p}_\text{min}=(2\pi/L)\vec{b}_1=(2\pi/L,0)$ and $\vec{p}_\text{min}=(2\pi/L)\vec{b}_2=(0,2\pi/L)$. For these momenta, $\Delta(\vec{p})$ takes the value
\begin{equation}
\Delta(\vec{p}_\text{min}) = 4\sin(\pi/L)^2.
\end{equation}

For a triangular lattice, the function $\Delta(\vec{p})$ is reported in Appendix~A of Ref.~\cite{PhysRevD.54.1782}:
\begin{equation}
\Delta(\vec{p}) = 4\left[1-\frac{1}{3}\left(\cos(p_x)+2\cos\left(\frac{p_x}{2}\right)\cos\left(\frac{\sqrt{3}p_y}{2}\right)\right)\right].
\label{Deltatri}
\end{equation}
The direct and reciprocal bases of the triangular lattice are
\begin{equation}
\vec{a}_1=\begin{pmatrix}1 \\ 0\end{pmatrix}, \ \vec{a}_2=\begin{pmatrix}\frac{1}{2} \\ \frac{\sqrt{3}}{2}\end{pmatrix}, \quad \vec{b}_1=\begin{pmatrix}1 \\ -\frac{1}{\sqrt{3}}\end{pmatrix}, \ \vec{b}_2=\begin{pmatrix}0 \\ \frac{2}{\sqrt{3}}\end{pmatrix},
\label{triangularbasis}
\end{equation}
with the same normalization as in Eq.~(\ref{basisnorm}). On a finite lattice with size $L$, the two minimum momenta are $\vec{p}_\text{min}=(2\pi/L)\vec{b}_1=(2\pi/L,-2\pi/\sqrt{3}/L)$ and $\vec{p}_\text{min}=(2\pi/L)\vec{b}_2=(0,4\pi/\sqrt{3}/L)$. For these momenta, $\Delta(\vec{p})$ takes the value
\begin{equation}
\Delta(\vec{p}_\text{min}) = \frac{16}{3}\sin(\pi/L)^2.
\label{Deltaminvalue}
\end{equation}
The fact that $\Delta(\vec{p})$ takes the same value for the two minimum momenta 
for both lattices considered here is a direct consequence of the invariance
under the symmetry of Eq.~(\ref{rotationsym}) with $\theta=\pi/2$ for the
square lattice, and $\theta=2\pi/3$ for the triangular lattice.

\subsection{Honeycomb lattice}\label{sec:xi:honeycomb}

\begin{figure}
\includegraphics[width=0.4\linewidth,keepaspectratio]{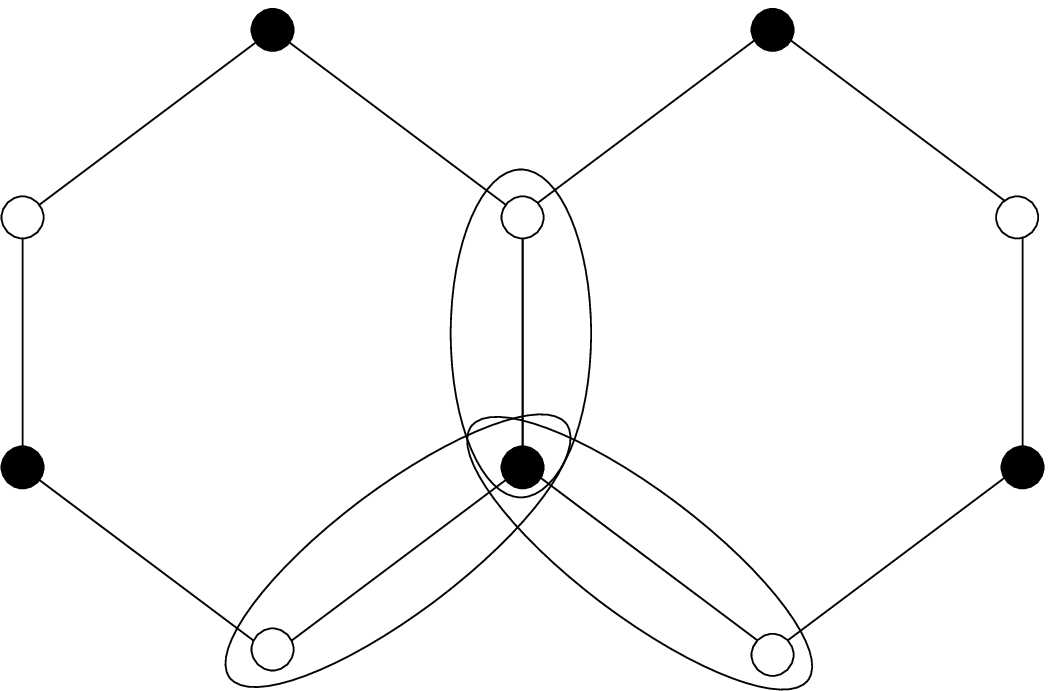}
\caption{A portion of a honeycomb lattice, which can be
  considered as a triangular lattice with a unit cell of two sites. The
  filled (empty) circles are sites on the $A$ ($B$) sublattice. The ellipses indicate three possible choices for the unit cell. Rotations by $\theta=2\pi/3$ map the possible choices for the unit cell onto each other.}
\label{hexagon}
\end{figure}

Since the honeycomb lattice can be considered as a triangular lattice where the elementary cell has two sites, the two-point function $\C(\vec{x})$ of a local order parameter constructed on a single elementary unit cell can be defined so that its domain is a triangular lattice, i.e., $\vec{x}=n_1\vec{a}_1+n_2\vec{a}_2$, with the lattice basis $\{\vec{a}_1,\vec{a}_2\}$ given in Eq.~(\ref{triangularbasis}).
However, different than in the case of a triangular lattice, the two-point
function $\C(\vec{x})$ may not be invariant under the rotation of
Eq.~(\ref{rotationsym}) with $\theta=2\pi/3$. In fact, such a symmetry holds
for some choices of the order parameter only. If the local order parameter
$\phi(\vec{x})$ in the unit cell $\vec{x}$ is defined in terms of observables
at lattice site $\vec{x}_A$ ($\vec{x}_B$) that belongs to the $A$ ($B$)
sublattice, then effectively the two-point function $\C(\vec{x})$ is
invariant under the rotation group of the triangular lattice, i.e., the
rotation of Eq.~(\ref{rotationsym}) with $\theta=2\pi/3$. For instance, this
is the case when the order parameter is the $A$ or $B$ sublattice
magnetization. In this work, we have considered the antiferromagnetic order
parameters given in Eqs.~(\ref{ophubbard}) and (\ref{opkm}). For these
local order parameters, which involve a combination of the $A$ and $B$
sublattice magnetization, the two-point function $\C(\vec{x})$ is not
invariant under a rotation by $\theta=2\pi/3$. The reason lies in the ambiguity in defining the elementary unit cell of the honeycomb lattice. As illustrated in Fig.~\ref{hexagon}, there are three possible choices for defining the elementary unit cell; a rotation by $\theta=2\pi/3$ maps one possible unit cell to another.

The absence of the lattice rotational symmetry for $\C(\vec{x})$ requires a
generalization of the arguments given in Appendix~\ref{sec:xi:regular}. To
this end, let us consider in full generality a finite lattice that extends
over $L_1$ ($L_2$) lattice unit cells in the direction parallel to $\vec{a}_1$
($\vec{a}_2$). For such a lattice, there are two minimum momenta
\begin{align}
\vec{p}^{(1)}_\text{min}&=\frac{2\pi}{L_1}\vec{b}_1=\left(\frac{2\pi}{L_1},-\frac{2\pi}{\sqrt{3}L_1}\right)\,,\\
\vec{p}^{(2)}_\text{min}&=\frac{2\pi}{L_2}\vec{b}_2=\left(0,\frac{4\pi}{\sqrt{3}L_2}\right)\,. 
\end{align}
A straight-forward generalization of Eq.~(\ref{xi2finite}) consists in defining
a finite-size correlation length $\xi^{(i)}(L)$ for each principal direction
$i=1,2$ as
\begin{equation}
\xi^{(i)}(L)^2 \equiv \frac{1}{\Delta(\vec{p}^{(i)}_\text{min})}\left[\frac{\widetilde{\C}(0)}{\widetilde{\C}(\vec{p}^{(i)}_\text{min})}-1\right], \quad i=1,2,
\label{xi2finiteani}
\end{equation}
where $\Delta(\vec{p})$ is given in Eq.~(\ref{Deltatri}). Even if, due to the
lack of the lattice rotational symmetry,
$\widetilde{\C}(\vec{p}^{(1)}_\text{min})\ne
\widetilde{\C}(\vec{p}^{(2)}_\text{min})$, for $L_1=L_2=L$ it is possible to
define an averaged correlation length by taking the mean value of
$\widetilde{\C}(\vec{p})$ over the two minimum momenta: 
\begin{equation}
\xi(L)^2 \equiv \frac{1}{\Delta(\vec{p}_\text{min})}\left[\frac{\widetilde{\C}(0)}{\left(\widetilde{\C}(\vec{p}^{(1)}_\text{min})+\widetilde{\C}(\vec{p}^{(2)}_\text{min})\right)/2}-1\right].
\label{xi2finiteave}
\end{equation}
For $L_1=L_2=L$, $\Delta(\vec{p})$ takes the same value given in
Eq.~(\ref{Deltaminvalue}) at the two minimum momenta $\vec{p}^{(1)}_\text{min}$ and $\vec{p}^{(2)}_\text{min}$ (see the discussion at end of Appendix~\ref{sec:xi:regular}). The definition of $\xi(L)$ given in Eq.~(\ref{xi2finiteave}) corresponds to a generalized $f$-mean value of $\xi^{(1)}(L)$ and $\xi^{(2)}(L)$,
\begin{equation}
\xi(L) = f^{-1}\left(\frac{f(\xi^{(1)}(L))+f(\xi^{(2)}(L))}{2}\right)\,,
\end{equation}
where $f(x)$ is a monotonic positive function
\begin{equation}
f(x)=\frac{1}{1+x^2\Delta(\vec{p}_\text{min})}.
\end{equation}
Moreover, if $\xi^{(i)}/L$ are RG-invariant quantities, then $\xi/L$ is
also an RG-invariant observable.

\subsection{Correlation length from real-space correlations}\label{sec:xi:real}

An alternative definition of the finite-size correlation length can be
obtained by directly considering Eq.~(\ref{xi2def}) and extending the sum
over the (finite) set of lattice sites. With periodic boundary conditions, such a
prescription does not uniquely fix the definition of $\xi$. To be specific,
as in Appendix~\ref{sec:xi:honeycomb}, we consider a finite lattice that
extends over $L_i$ lattice sites in the direction parallel to $\vec{a}_i$, with
$i=1,2$. With periodic boundary conditions, the two-point function satisfies
$\C(\vec{x})=\C(\vec{x}+nL_1\vec{a}_1+mL_2\vec{a}_2)$ for arbitrary integers $n$ and $m$. However, the Euclidean length $|\vec{x}|$ in
Eq.~(\ref{xi2def}) is not invariant under translations. This
leaves us the freedom to define the correlation length as a sum over
$\vec{x}=n_1\vec{a}_1+n_2\vec{a}_1$, where $n_i$ runs over
$-(L_i-1)+l_i,-(L_i-1)+l_i+1,\ldots,l_i-1,l_i$, with arbitrary $l_i$. In
order to have a nontrivial FSS limit, the maximum value of the index $l_i$
must be proportional to $L_i$.

These considerations lead us to define a finite-size correlation length $\xi_{s,\kappa,\rho}(L)$ as
\begin{equation}
\xi_{s,\kappa,\rho}(L)^2 \equiv
\frac{\smashoperator[r]{\sum_{\substack{(-1+\kappa)L_1+1\le n_1\le \kappa L_1 \\(-1+\rho)L_2+1\le n_2\le \rho L_2}}}|n_1\vec{a}_1+n_2\vec{a}_2|^2\C(n_1\vec{a}_1+n_2\vec{a}_2)}{\smashoperator{\sum_{\substack{0\le n_1\le L_1-1 \\0\le n_2\le L_2-1}}}\C(n_1\vec{a}_1+n_2\vec{a}_2)}.
\label{xi2finites}
\end{equation}
We note that, by virtue of the aforementioned translational invariance, in the denominator of Eq.~(\ref{xi2finites}) a shift of the sum as done for the numerator does not change the result. In Eq.~(\ref{xi2finites}), the choice of $\kappa=\rho=1/2$ corresponds to defining the distance $|\vec{x}|$ as the minimum one.

Although in the infinite-volume limit
$L_1,L_2\rightarrow\infty$ {\it at fixed $U$} the correlation lengths as
defined in Eqs.~(\ref{xi2finites}) and (\ref{xi2finite}) converge to the
same observable, in the FSS limit these definitions of $\xi$, as well as
those given in Eqs.~(\ref{xi2finiteani}) and (\ref{xi2finiteave}),
correspond to different observables. As a consequence, the corresponding
ratios $\xi/L$ constructed with the various definitions of $\xi$ [see
Eqs.~(\ref{Rxi1-def})\,--\,(\ref{Rxispi-def})] correspond to different
RG-invariant quantities. This in particular affects the corrections to
scaling which, as shown in Sec.~\ref{sec:results}, can be significantly
different. In particular, setting $\kappa=\rho=0$ in Eq.~(\ref{xi2finites})
gives rise to a large contribution of the numerator when $|\vec{x}|\approx
L_1,L_2$ because, for such values of $\vec{x}$ and due to the periodic
boundary conditions, $\C(\vec{x})\approx \C(0)$. This results in a large
background term due to the nonuniversal short-distance part of the
correlation function that gives rise to large corrections to scaling.

Finally, we observe that Eq.~(\ref{xi2finites}) is correctly defined only
when $\kappa L_1$ and $\rho L_2$ are integer numbers. In order to be able to
extrapolate to the FSS limit, this property must hold for every lattice
size. Such limitations on the values of $\kappa$ and $\rho$, together with
the limitations on the lattice sizes that can be simulated (see
Sec.~\ref{sec:method}), further limit the applicability of
Eq.~(\ref{xi2finites}) for generic values of $\kappa$ and $\rho$. For the
honeycomb Hubbard and the Kane-Mele-Hubbard models we simulated lattices
with $L_1=L_2=L$, with $L$ being a multiple of $3$. For this reason, we
employed the definition in Eq.~(\ref{xi2finites}) with $\kappa=\rho=1/3$. In
the case of the $\pi$-flux Hubbard model, we simulated lattices with
$L_1=L/2$ and $L_2=L$, with $L$ being a multiple of $4$. This leads us to
either choose $\kappa=\rho=1/2$ or $\kappa=1/2$ and $\rho=1/4$, the latter
giving rise to smaller scaling corrections (see
Sec.~\ref{sec:results:piflux}).

\section{Finite-size scaling analysis of $\chicl$ at fixed $U$ for the honeycomb Hubbard model}
\label{sec:fsschicl}
\begin{table}
\caption{Results of the fit of $\chicl$ for the honeycomb Hubbard model to
  Eq.~(\ref{chi_vs_U_Taylor}) (first three sets) and to
  Eq.~(\ref{chi_vs_U_Taylor_background}) (last two sets), for $U\in
  [3.6,4]$. The critical exponent $\eta'$ is defined as
  $\eta'\equiv\eta+z-1$, with $\eta'=\eta$ if $z=1$. $L_{\rm min}$ is
  the minimum lattice size taken into account in the fits. In the quoted
  error bars for $\eta'$, the first number reports the statistical precision
  as obtained from the fit, while the second number gives the sum of the
  maximum variation in the results upon varying $U_c$ and upon varying $\nu$
  within one error bar, as quoted in
  Eqs.~(\ref{Uc_hubbard}) and (\ref{nu_hubbard}). The corresponding maximum
  oscillation of $\chi^2$ is reported between parentheses after its central
  value.}
\begin{ruledtabular}
\begin{tabular}{r@{\hspace{1em}}l@{}@{}.l@{}.@{}r}
& \multicolumn{1}{c}{$L_{\rm min}$} & \multicolumn{1}{c}{$\eta'$} & \multicolumn{1}{c}{$\chi^2/\DOF$} \\
\hline
\multirow{3}{*}{$n_{\rm max}=1$} & $6$  & 0.7154(8+79) &  $2832(985)/22$ \\
& $9$  & 0.696(1+11)  &  $1902(481)/17$ \\
& $12$ & 0.671(3+14)  &   $894(107)/12$ \\
\\
\multirow{3}{*}{$n_{\rm max}=2$} & $6$  & 0.7359(9+94) &   $644(519)/21$ \\
& $9$  & 0.735(2+12)  &   $383(271)/16$ \\
& $12$ & 0.731(4+13)  &    $110(67)/11$ \\
\\
\multirow{3}{*}{$n_{\rm max}=3$} & $6$  & 0.7324(9+85) &    $213(167)/20$ \\
& $9$  & 0.731(2+11)  &     $129(77)/15$ \\
& $12$ & 0.734(4+15)  &      $45(20)/10$ \\
\\
$n_{\rm max}=2$ & $6$  & 0.887(7+72)  &   $142(87)/20$ \\
$m_{\rm max}=0$ & $9$  & 0.93(1+8)    &   $24.2(10.3)/15$ \\
\\
$n_{\rm max}=2$ & $6$  & 0.78(1+6)    &   $84(29)/19$ \\
$m_{\rm max}=1$ & $9$  & 0.83(5+7)    &   $19(6)/14$ \\
\\
$n_{\rm max}=2$ & $6$  & 0.79(2+5)    &   $80(31)/18$ \\
$m_{\rm max}=2$ & $9$  & 0.79(5+6)    &   $17.3(3.4)/13$ \\
\end{tabular}
\end{ruledtabular}
\label{fit_chi_vs_U_hubbard}
\end{table}

In order to further assess the reliability of the results of Sec.~\ref{sec:results:hubbard} and the
overall consistency of the estimates of the critical exponents for the honeycomb Hubbard model, we
analyzed the FSS behavior of $\chicl$ as a function of $U$ and $L$, as we
did for the RG-invariant quantity $\Rxist$. To this end, we consider a
Taylor expansion of the right-hand side of Eq.~(\ref{chifss}). Neglecting scaling
corrections, we fit our data for $\chicl$ to
\begin{equation}
\chicl(U,L)=L^{1-\eta'}\sum_{n=0}^{n_{\rm max}}a_n(U-U_c)^nL^{n/\nu},
\label{chi_vs_U_Taylor}
\end{equation}
leaving $\eta'$, $\{a_n\}$ as free parameters, and using the values of $U_c$
and $\nu$ as given by Eqs.~(\ref{Uc_hubbard}) and (\ref{nu_hubbard}). We repeat
the fit by varying $U_c$ and $\nu$ within one error bar as quoted in
Eqs.~(\ref{Uc_hubbard}) and (\ref{nu_hubbard}). As in the FSS analysis of
$\Rxist$, we restrict the analysis to values $U\in [3.6,4]$ and
systematically disregard the smallest lattice sizes. The fit results are
reported in Table~\ref{fit_chi_vs_U_hubbard}. Inspection of the results
reveals a significant decrease of the $\chi^2/\DOF$ ratio when we increase
$n_{\rm max}$ from $n_{\rm max}=1$ to $n_{\rm max}=2$, and a smaller decrease in
$\chi^2/\DOF$ when $n_{\rm max}$ is further increased to $n_{\rm
  max}=3$. Such a decrease in the $\chi^2/\DOF$ ratio is even less
statistically relevant if we take into account the oscillations in the value
of $\chi^2/\DOF$ due to the uncertainty in $U_c$ and $\nu$. Moreover, the
fitted values for $n_{\rm max}=2$ and $n_{\rm max}=3$ are in agreement with
each other, suggesting that within the statistical accuracy a Taylor
expansion with $n_{\rm max}=2$ is sufficient to describe the data. We also
observe that the main contribution to the error bars is due to the
uncertainty in $U_c$ and $\nu$.

In line with the findings of Table~\ref{fit_chi_vs_R_hubbard}, even
considering the maximum oscillation of $\chi^2/\DOF$ upon variation of $U_c$ and
$\nu$ within one error bar as quoted in
Eqs.~(\ref{Uc_hubbard}) and (\ref{nu_hubbard}), all of the fits have a large
$\chi^2/\DOF$. This confirms the importance of scaling corrections. To monitor
their role, we repeat the fits including a scaling correction in the form of
a background term, [see Eq.~(\ref{chi_vs_R_Taylor_background})]. To this end, we use
\vspace{-0.08em}
\begin{equation}
\begin{split}
\chicl(U,L)=L^{1-\eta'}&\sum_{n=0}^{n_{\rm max}}a_n(U-U_c)^nL^{n/\nu} \\
+&\sum_{m=0}^{m_{\rm max}}b_m(U-U_c)^m.
\end{split}
\label{chi_vs_U_Taylor_background}
\end{equation}
Fit results for $n_{\rm max}=2$ and three values of $m_{\rm max}$ are shown
in Table~\ref{fit_chi_vs_U_hubbard}. Upon increasing $m_{\rm max}$ from $m_{\rm max}=0$ to
$m_{\rm max}=1$, we observe a decrease in the $\chi^2/\DOF$ ratio that is,
however, less significant if we consider the oscillation in the value of
$\chi^2/\DOF$ due to the uncertainty in $U_c$ and $\nu$. A further increase of
$m_{\rm max}$ to $m_{\rm max}=2$ does not significantly change the
$\chi^2/\DOF$ ratio. Accordingly, the expansion with $n_{\rm max}=2$, $m_{\rm max}=1$ should adequately describe the data. The corresponding fits exhibit
a small $\chi^2/\DOF$ for $L_{\rm min}=9$, and the resulting value of
$\eta'=0.83(12)$ is in agreement with the estimate of
Eq.~(\ref{eta_hubbard}). Moreover, this value agrees with the fit for $L_{\rm min}=6$, and
also with the fits obtained by setting $n_{\rm max}=m_{\rm max}=2$.

\bibliographystyle{apsrev4-1_custom}
\bibliography{refs2,fassaad2}
\end{document}